\documentclass[twocolumn]{aastex61}
\hypersetup{colorlinks,linkcolor={red!60!black},citecolor={blue!60!black},urlcolor={blue!60!black}}

\usepackage{graphicx}	
\usepackage{amsmath}	
\usepackage{amssymb}	
\usepackage[para,flushleft]{threeparttable} 
\usepackage{epstopdf}
\usepackage{textgreek}

\usepackage{natbib,twoopt}
\bibpunct{(}{)}{;}{a}{}{,} 
\makeatletter
\newcommandtwoopt{\citeads}[3][][]{\href{http://adsabs.harvard.edu/abs/#3}%
{\def\hyper@linkstart##1##2{}%
\let\hyper@linkend\@empty\citealp[#1][#2]{#3}}}
\newcommandtwoopt{\citepads}[3][][]{\href{http://adsabs.harvard.edu/abs/#3}%
{\def\hyper@linkstart##1##2{}%
\let\hyper@linkend\@empty\citep[#1][#2]{#3}}}
\newcommandtwoopt{\citetads}[3][][]{\href{http://adsabs.harvard.edu/abs/#3}%
{\def\hyper@linkstart##1##2{}%
\let\hyper@linkend\@empty\citet[#1][#2]{#3}}}
\newcommandtwoopt{\citeyearads}[3][][]%
{\href{http://adsabs.harvard.edu/abs/#3}
{\def\hyper@linkstart##1##2{}%
\let\hyper@linkend\@empty\citeyear[#1][#2]{#3}}}
\makeatother

\shorttitle{Suppression of CMEs by an overlying large-scale magnetic field}
\shortauthors{Alvarado-G\'omez et~al.}

\pdfoutput=1 
\usepackage{amsmath,amstext}
\usepackage[T1]{fontenc}
\usepackage[figure,figure*]{hypcap}

\begin{document}

\title{Suppression of Coronal Mass Ejections in active stars by an overlying large-scale magnetic field: A numerical study}

\correspondingauthor{Juli\'an David Alvarado G\'omez}
\email{julian.alvarado-gomez@cfa.harvard.edu}

\author[0000-0001-5052-3473]{Juli\'an~D.~Alvarado-G\'omez}
\altaffiliation{\href{https://www.cfa.harvard.edu/~jalvarad/}{AstroRaikoh}}
\affiliation{Harvard-Smithsonian Center for Astrophysics, 60 Garden St. Cambridge, MA 02138, USA}
\affiliation{European Southern Observatory, Karl-Schwarzschild-Str. 2, 85748 Garching bei M\"unchen, Germany}

\author{Jeremy~J.~Drake}
\affiliation{Harvard-Smithsonian Center for Astrophysics, 60 Garden St. Cambridge, MA 02138, USA}

\author{Ofer~Cohen}
\affiliation{University of Massachusetts at Lowell, Department of Physics \& Applied Physics, 600 Suffolk St., Lowell, MA 01854, USA}

\author{Sofia P. Moschou}
\affiliation{Harvard-Smithsonian Center for Astrophysics, 60 Garden St. Cambridge, MA 02138, USA}

\author{Cecilia Garraffo}
\affiliation{Harvard-Smithsonian Center for Astrophysics, 60 Garden St. Cambridge, MA 02138, USA}




\begin{abstract}
\noindent We present results from a set of numerical simulations aimed at exploring the mechanism of coronal mass ejection (CME) suppression in active stars by an overlying large-scale magnetic field. We use a state-of-the-art 3D magnetohydrodynamic (MHD) code which considers a self-consistent coupling between an Alfv\'en wave-driven stellar wind solution, and a first-principles CME model based on the eruption of a flux-rope anchored to a mixed polarity region. By replicating the driving conditions used in simulations of strong solar CMEs, we show that a large-scale dipolar magnetic field of $75$~G is able to fully confine eruptions within the stellar corona. Our simulations also consider CMEs exceeding the magnetic energy used in solar studies, which are able to escape the large-scale magnetic field confinement. The analysis includes a qualitative and quantitative description of the simulated CMEs and their dynamics, which reveals a drastic reduction of the radial speed caused by the overlying magnetic field. With the aid of recent observational studies, we place our numerical results in the context of solar and stellar flaring events. In this way, we find that this particular large-scale magnetic field configuration establishes a suppression threshold around $\sim$\,$3 \times 10^{32}$~erg in the CME kinetic energy. Extending the solar flare-CME relations to other stars, such CME kinetic energies could be typically achieved during erupting flaring events with total energies larger than $6 \times 10^{32}$~erg (GOES class $\sim$X70).
  
\end{abstract}

\keywords{magnetohydrodynamics (MHD) --- methods: numerical --- stars: activity --- stars: magnetic field --- stars: winds, outflows --- Sun: coronal mass ejections (CMEs)}



\section{Introduction} \label{sec:intro}

\noindent Flares and coronal mass ejections (CMEs) constitute the most energetic phenomena in the solar system. Succinctly, the former correspond to a sudden radiation flash in the solar atmosphere (covering the entire electromagnetic spectrum), and the latter to a relatively localized release of magnetized plasma into interplanetary space. The standard picture of their generation involves a chain of processes which is ultimately linked to the re-organization or reconnection of the coronal magnetic field (see \citeads{2011LRSP....8....1C}, \citeads{2011LRSP....8....6S}, \citeads{2012LRSP....9....3W}, \citeads{2017LRSP...14....2B}). 

Recent correlation studies have shown that the most energetic solar flares (X-class in the \textit{Geostationary Operational Environmental Satellite -- GOES\footnote[1]{\url{http://www.swpc.noaa.gov/products/goes-x-ray-flux}}} -- scale, which is based on the flare peak emission in the SXR band of $1$\,$-$\,$8$~\AA), are nearly always accompanied by a CME (\citeads{2003SoPh..218..261A}, \citeads{2009IAUS..257..233Y}, \citeads{2017SoPh..292....5C}). On the other hand, while the complete energy budget is still uncertain, it is nowadays clear that the CMEs carry a much larger fraction of the available energy than the X-ray/EUV flare counterparts (see Emslie et al. \citeyearads{2005JGRA..11011103E}, \citeyearads{2012ApJ...759...71E}). Still, the flare-CME energies seem to reach comparable levels when thermal and non-thermal emission in other wavelengths are included (i.e.~white light/UV continuum; \citeads{2010NatPh...6..690K}, \citeads{2011A&A...530A..84K}, \citeads{2012ApJ...759...71E}, \citeads{2017ApJ...836...17A}).  

While permanent monitoring of the Sun has driven significant progress in this field, some observational aspects seem to indicate that the flare-CME paradigm deviates from its solar rendition in the stellar case. This includes the fact that the coronae of active stars appear to be continuously flaring (e.g., \citeads{2002ApJ...580.1118K}, \citeads{2010ApJ...723.1558H}), and their light curves can often be well-described using a superposition of flares (c.f.,~\citeads{2000ApJ...541..396A}, \citeads{2007A&A...471..645C}). At face value this enhanced flare activity would imply a correspondingly high occurrence of stellar CMEs, possibility that has been considered by different authors in the context of stellar evolution (c.f., \citeads{2012ApJ...760....9A}, \citeads{2013ApJ...764..170D}, \citeads{2015MNRAS.448.1628S}, \citeads{2017ApJ...840..114C}) and environmental conditions --such as habitability-- around stellar systems (c.f., \citeads{2007AsBio...7..167K}, \citeads{2007AsBio...7..185L}, \citeads{2011ApJ...738..166C}, \citeads{2016ApJ...826..195K}, \citeads{2016ApJ...833L...8T}, \citeads{2017ApJ...846...31C}). Flares in very active stars can be $10^3 - 10^6$ times more energetic than solar flares (c.f. \citeads{2007ApJ...654.1052O}, \citeads{2010ApJ...714L..98K}, \citeads{2013ApJS..209....5S}, \citeads{2016ApJ...829...23D}). Despite this, convincing evidence for stellar CMEs is rare and elusive.  A handful of studies have reported blue-wing enhancements in the Balmer lines of M dwarfs at velocities exceeding the stellar escape velocity which the authors interpreted as eruptive stellar filaments (\citeads{1990A&A...238..249H}, \citeads{1994A&A...285..489G}, \citeads{2016A&A...590A..11V}). \citetads{1997A&A...321..803G} found a distinct H$\alpha$ blue-wing enhancement in a spectrum of the weak T Tauri star DZ Cha (RX J1149.8--7850) they attributed to a CME and deduced an ejected mass in the range of $10^{18} - 10^{19}$~g, which is a factor $10-100$ larger than for the most massive solar CMEs. \citetads{2017ApJ...850..191M} followed the suggestion of \citetads{1999A&A...350..900F}, that an absorption event during a giant flare on Algol could be a CME, to deduce an even larger ejected mass in the range $10^{21} - 10^{22}$~g. However, several studies attempting to detect of stellar CMEs using radio and optical data found no significant signatures (e.g., \citeads{2014MNRAS.443..898L}, \citeads{2016ApJ...830...24C}, \citeads{2017PhDT.........8V}, \citeads{2018ApJ...856...39C}). 

In addition, as discussed in detail by \citetads{2013ApJ...764..170D}, problematic consequences arise in extrapolating solar flare-CME relations to active stars. One of them is a very large predicted contribution from CMEs to the stellar mass loss rate ($\dot{M}$), leading to $\dot{M}$ values up to four orders of magnitude greater that the accepted value for the Sun (due to the solar wind), and roughly two orders of magnitude above the largest $\dot{M}$ estimate obtained from astrospheric Lyman-$\alpha$ absorption (Wood~et~al.~\citeyearads{2002ApJ...574..412W}, \citeyearads{2005ApJ...628L.143W}, \citeyearads{2014ApJ...781L..33W}). In a previous study following a similar extrapolation procedure, \citetads{2012ApJ...760....9A} obtained comparable $\dot{M}$ values in the case of T~Tauri stars. 

 Likewise, \citetads{2015ApJ...809...79O} obtained similar $\dot{M}$ predictions by extending flare-CME relationships to include the different energy band passes and flare frequencies observed in solar-type and low-mass stars. Following a slightly different approach, \citeads{2017ApJ...840..114C} considered a solar-motivated relation between surface-averaged magnetic fluxes and the contribution of CMEs to the total mass loss in stars. In that model, the CMEs would dominate by factors of $10-100$ the mass loss budget in young stars (few Myr in age), surpassing the contribution from the steady stellar wind even for later stages (younger than about 1 Gyr). Recently, \citetads{2017MNRAS.472..876O} presented an empirical model (which also incorporates solar extrapolations), with relatively good agreement on the predicted $\dot{M}$ for moderately active stars (i.e., surface X-ray fluxes $F_{\rm X} < 10^{6}$~erg~cm$^{-2}$~s$^{-1}$), but still extremely high values at the high-activity end.

The analysis of \citetads{2013ApJ...764..170D} also showed that in order to sustain solar-like flare-driven CME activity, the kinetic energy requirements would be implausibly high, reaching up to ten per cent of the stellar bolometric luminosity for active stars in the saturated regime. They concluded that either the relationships between CME mass/speed and flare energy must flatten for X-ray energies $\gtrsim 10^{31}$ erg, or the flare-CME association rate must drop significantly below 1 for more energetic events (unlike the solar case). 

One process that could reduce the flare-CME association rate in active stars, avoiding the energy/mass loss quandary identified by \citetads{2013ApJ...764..170D}, is the suppression of the plasma ejecta by a strong overlying magnetic field  (\citeads{2016IAUS..320..196D}; see also \citeads{2017MNRAS.472..876O}). This possibility has been invoked to explain the behaviour of certain flare-rich CME-less active regions in the Sun (e.g. Thalmann~et~al.~\citeyearads{2015ApJ...801L..23T}, \citeyearads{2017ApJ...844L..27T}, \citeads{2015ApJ...804L..28S}, \citeads{2016ApJ...826..119L}). In the case of active stars, there is ample evidence of strong large-scale magnetic fields that could provide the required confining conditions (see \citeads{2009ARA&A..47..333D}, \citeads{2011IAUS..271...23D}, \citeads{2017NatAs...1E.184S}). Such a scenario was briefly illustrated in the magneto-hydrodynamic (MHD) simulation presented by \citetads{2016IAUS..320..196D}, where a strong solar CME failed to erupt when placed under the large-scale magnetic field configuration of a much more active star (AB~Doradus; \citeads{2002ApJ...575.1078H}). 

In this paper we follow a similar methodology, using a set of realistic simulations of magnetically-driven CMEs embedded in a large-scale magnetic field, in order to determine their properties and ability to escape the confining field. This study represents the first step in characterizing the conditions and regimes of operation of this mechanism. Section~\ref{sec_models} contains the description of the numerical models, as well as the different boundary and initial conditions. Our results are presented in Sect.~\ref{sec_results} and discussed in the solar and stellar context in Sect.~\ref{sec_discussion}. A summary and the conclusions of our work are presented in Sect.~\ref{sec_summary}. 

\section{Numerical Simulations}\label{sec_models}

\noindent Two different coupled models are considered in this study: the Alfv\'en Wave Solar Model (AWSoM, \citeads{2013ApJ...764...23S}, \citeads{2014ApJ...782...81V}, \citeads{2016arXiv160904379S}), and the Gibson \& Low (GL) flux-rope CME model (\citeads{1998ApJ...493..460G}, \citeads{2004JGRA..109.1102M}, \citeads{2017ApJ...834..173J}). Both models are implemented in the latest version of the Space Weather Modeling Framework (SWMF, \citeads{2012JCoPh.231..870T}), and are now used extensively in the solar system context (e.g., \citeads{2015MNRAS.454.3697M}, \citeads{2017ApJ...834..172J}, \citeads{2017ApJ...845...98O}). 

The AWSoM model considers a non-ideal MHD regime, solving the equations for the conservation of mass, momentum, energy, and magnetic induction on a spherical grid. The coronal heating and stellar wind acceleration are calculated in a self-consistent manner from the propagation, reflection, and turbulent dissipation of Alfv\'en waves in the lower layers of the stellar atmosphere. These contributions are coupled in the form of an additional term for the total pressure in the momentum equation and a source term in the energy equation. Treatment of radiative losses and electron heat conduction are also considered in the simulation. Specific details of the numerical implementation can be found in \citetads{2014ApJ...782...81V} and \citetads{2016arXiv160904379S}.
  
Fixed values for the base temperature and density, matching expected solar chromospheric levels ($T_{0}~=~5~\times~10^{4}$~K and $n_{0} = 2 \times 10^{16}$ m$^{-3}$), are used as the inner boundary condition of the simulation. Likewise, the distribution of the magnetic field at the stellar surface over the course of one rotation period\footnote[2]{Known in the solar context as synoptic maps, i.e., averaged magnetograms over a solar Carrrington Rotation (CR).}, provides the boundary condition for calculating the initial configuration of the magnetic field in the three-dimensional domain. This is performed following a potential field extrapolation (\citeads{1969SoPh....9..131A}), under the numerical procedure described by \citetads{2011ApJ...732..102T}. 

The initial magnetic field configuration constitutes the main driver of the model, as it provides the directionality of the counter-propagating Alfv\'en waves (which follow the polarity of the field), and the amount of energy flow via the Poynting flux (which scales proportionally to the field strength at the surface). The simulation then evolves self-consistently until a steady-state solution is obtained for the stellar wind and corona. We have used this numerical approach in previous studies to characterize the corona and stellar wind environment around planet-hosting Sun-like stars (c.f.~Alvarado-G\'omez~et~al.~\citeyearads{2016A&A...588A..28A}, \citeyearads{2016A&A...594A..95A}), and M-dwarfs (c.f.~Garraffo~et~al.~\citeyearads{2016ApJ...833L...4G},~\citeyearads{2017ApJ...843L..33G}, \citeads{2017ApJ...834...14C}). For the simulations presented here we assumed solar values for the stellar mass, radius, and rotation period.   

\begin{figure}[!t]
\centering
\includegraphics[trim=0.1cm 0.1cm 1.6cm 0.3cm, clip=true,width=0.475\textwidth]{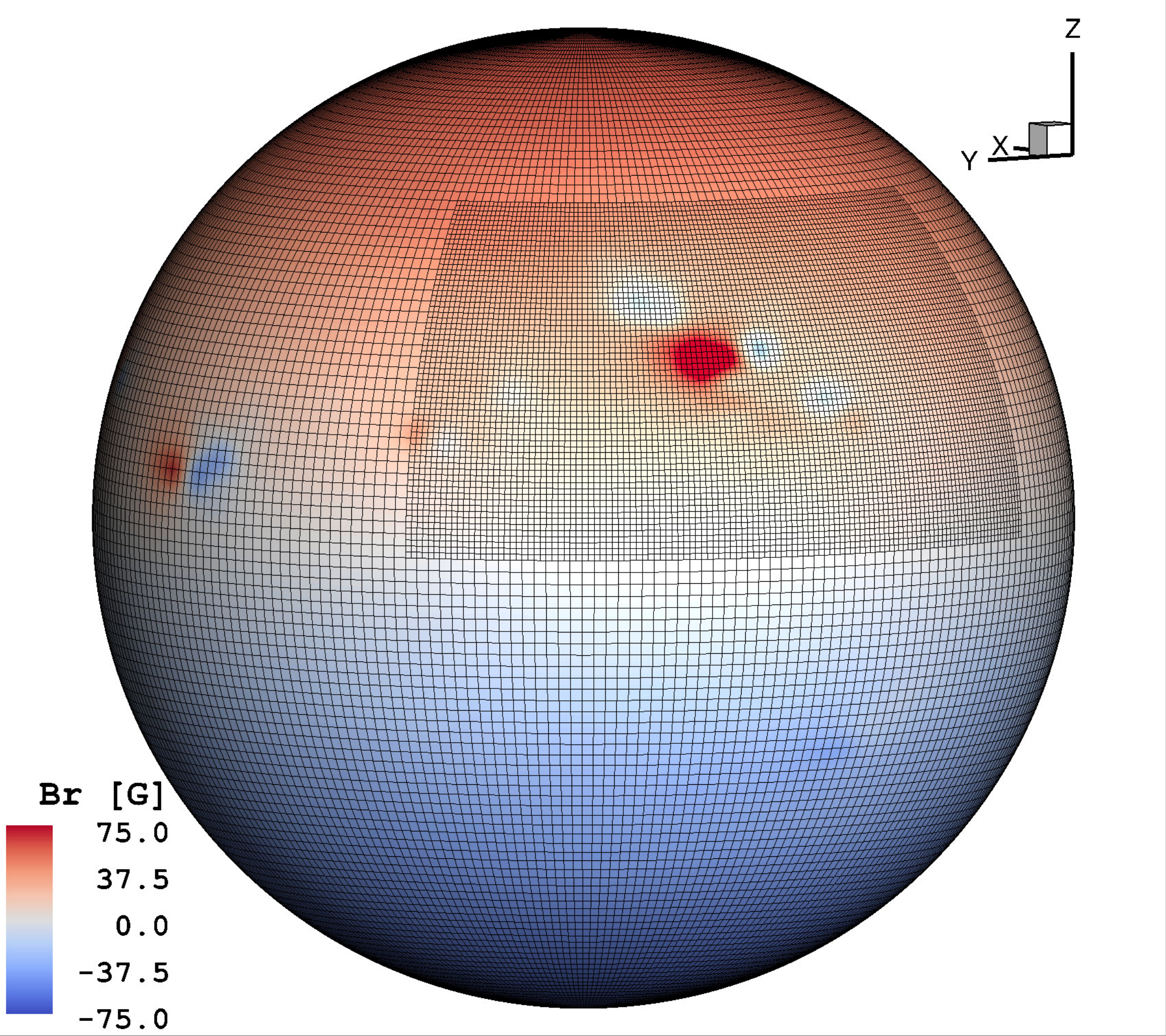}
\caption{Radial component of the magnetic field ($B_{\rm r}$) driving the simulation. Overlaid is the numerical grid at the stellar surface. The finer grid encloses the erupting active region and corresponds to the base of a high-resolution spherical wedge extending to half of the simulation domain.}
\label{fig_1}
\end{figure}

Since we are interested in the suppression of CMEs by an overlying field, the magnetic configuration driving our simulations is based on the superposition of two generic components: a large- and small-scale field (Fig.~\ref{fig_1}). For the first one we assumed a simple $75$ Gauss dipole aligned with the rotation axis of the star ($z$-axis). As reference, the dipole field strength of the Sun is on the order of $\sim1$ G in magnitude (\citeads{2008SoPh..252...19S}). The $75$ Gauss value was selected as a compromise between a sufficiently strong field to test the suppression mechanism, and the increased computational requirements resulting from this assumption\footnote[3]{Stronger magnetic fields not only require larger simulation domains for achieving a steady-state wind solution, but also significantly decrease the dynamic time-step for tracing the CME evolution, making the time-accurate runs highly demanding.}. However, as this is a modest large-scale field compared to reports from spectropolarimetric observations of very active stars (reaching up to kG levels; see \citeads{2009ARA&A..47..333D}, \citeads{2011IAUS..271...23D}), any confining effects on the CMEs could be potentially increased in those cases.  

For the small-scale component, we take one of the solar synoptic maps generated by the Global Oscillation Network Group (GONG) program\footnote[4]{\url{https://gong.nso.edu/data/magmap/}.}. While an arbitrary CR could have been considered (provided the presence of a prominent mixed polarity active region), we select the CR~2107 map because it was used for the calibration of the GL model (\citeads{2017ApJ...834..173J}), and for a numerical study of the propagation up to 1 AU of a CME emerging from the active region (AR) 11164 (\citeads{2017ApJ...834..172J}). We use this same active region to host the GL flux rope in our CME simulations. 

\begin{deluxetable}{lcc}[h!]
\tablecaption{Flux rope parameters initializing the GL CME simulations.\label{tab_1}}
\tablecolumns{3}
\tablenum{1}
\tablewidth{2pt}
\tablehead{\colhead{Parameter} & \colhead{Unit} & \colhead{Value}}
\startdata
Latitude & deg & 27.0 \\
Longitude & deg & 155.0 \\
Orientation\tablenotemark{a} & deg & 129.8 \\
Stretch ($a$) & \nodata & 0.6 \\
Pre-stretch distance ($r_1$) & $R_{*}$ & 1.8 \\
Size ($r_0$) &  $R_{*}$ & Variable \\
Magnetic strength ($a_1$) & G $R_*^{-2}$ & Variable\\
Flux rope helicity & \nodata & Dextral (+) \\
\enddata
\tablenotetext{a}{Measured with respect to the stellar equator in the clock-wise direction.}
\end{deluxetable}

We first obtain a steady-state corona and wind solution. The GL flux rope CME model is then linearly coupled inside the stellar corona domain. This model begins with a magneto-hydrostatic description of a twisted closed flux-rope, anchored to a mixed-polarity region of the magnetic field at the stellar surface (simulation's inner boundary). In addition to the magnetic flux rope, the GL model also embeds an initial density profile which follows the observed filament-cavity configuration typically preceding a CME event (\citeads{2013SoPh..284..179V}). Only the ambient coronal material is used to perform this density re-arrangement above the AR, so no mass is added by the GL model to the stellar corona domain. Due to the pressure imbalance of the structure with respect to the ambient stellar wind, the eruption process is not gradual but is immediately triggered once the simulation restarts. This is executed in time-accurate mode, following the evolution of the CME for a certain amount of time (depending on the problem of interest).

The starting state of the GL model is specified by a set of eight parameters involving the location, orientation, geometry, and magnetic properties of the erupting flux rope. These parameters connect the characteristics of the AR hosting the eruption, and the stretching transformation used to construct the GL model (see \citeads{1998ApJ...493..460G}). For direct comparison, our simulations use the same fixed parameters assumed in the solar calibration study, varying only the properties connected with the size ($r_0$) and magnetic strength ($a_1$) of the CME (Table~\ref{tab_1}; see also Fig.~2-a in \citeads{2017ApJ...834..173J}). As described by \citetads{2017ApJ...834..173J} and \citetads{2017JGRA..122.7979B}, this selection allows us to compute directly the poloidal flux of the erupting flux rope, $\Phi_{\rm p}$, as $\Phi_{\rm p} = c \cdot a_1 (r_0)^4$ with $c = 1.97 \times 10^{22}$. In this way, we perform 12 GL flux rope CME simulations with associated poloidal flux values in the range $10^{21}$ -- $10^{23}$ Mx.    

\begin{figure*}[!ht]
\centering
\includegraphics[width=0.495\textwidth]{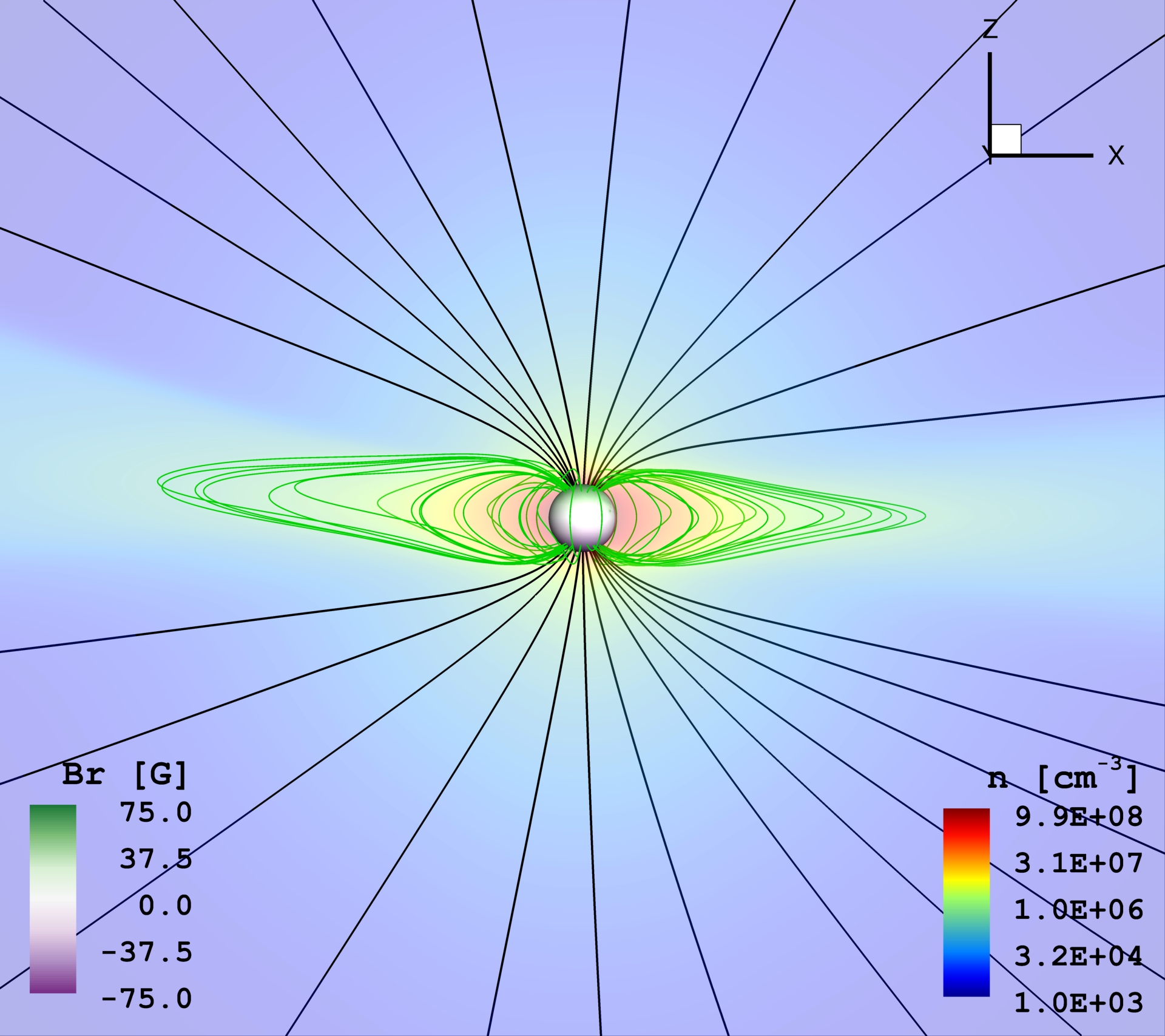}\hspace{1.2pt}\includegraphics[width=0.495\textwidth]{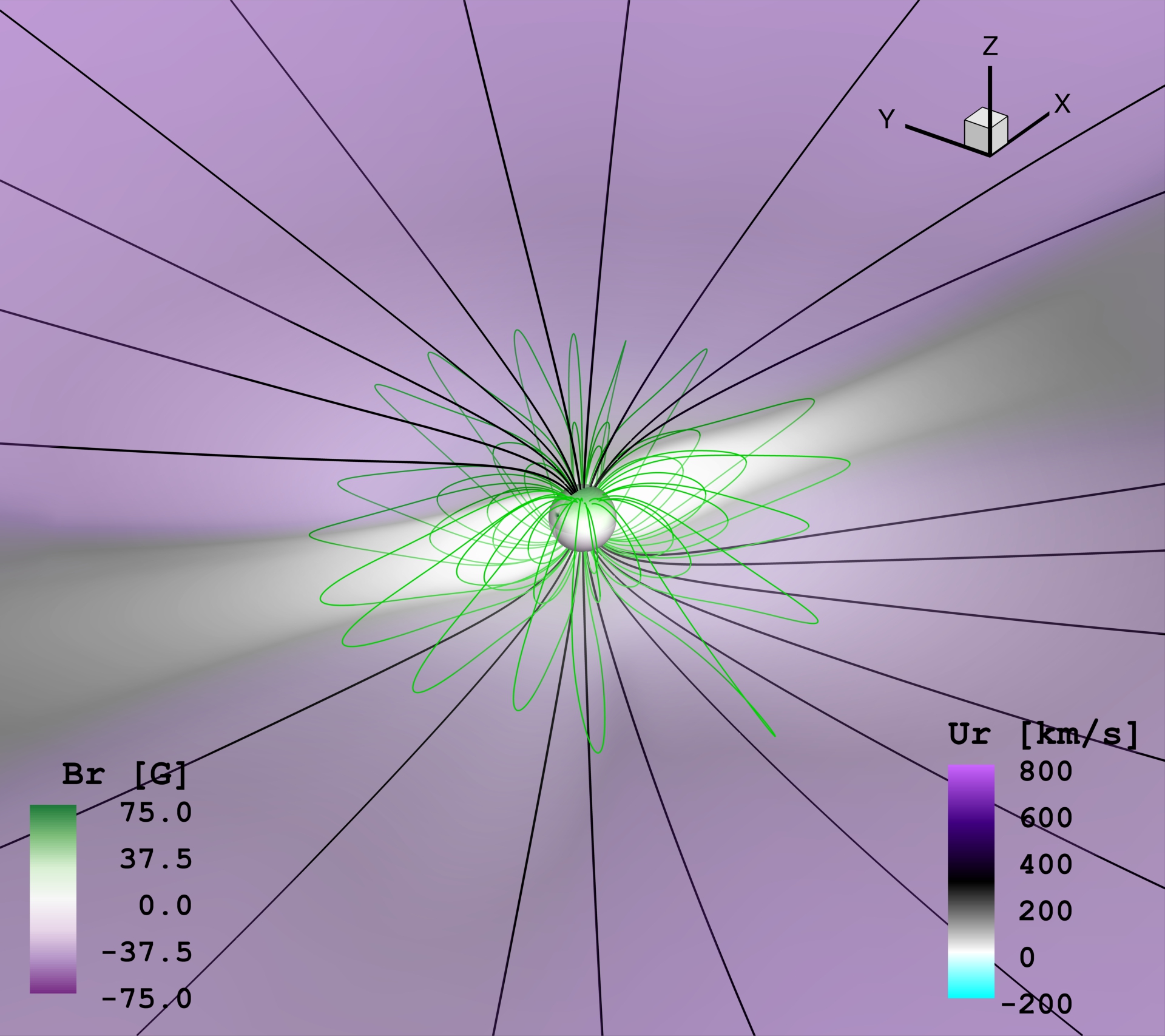}
\caption{Steady-state stellar wind solution obtained with the AWSoM model. The central sphere corresponds to the stellar surface, colored by the radial magnetic field configuration driving the simulation (Fig. \ref{fig_1}). The \textit{left panel} shows the distribution of the plasma density ($n$) over the meridional plane $y = 0$. The \textit{right panel} contains the resulting radial wind speed  ($U_{\rm r}$), projected onto a transversal plane aligned with the main polarity inversion line (PIL) of the eruptive AR (see Fig. \ref{fig_1}). Selected open and closed magnetic field lines are shown in black and green, respectively. The field of view (i.e.~side length of the visualization) in both panels is 36~$R_{*}$.}
\label{fig_2}
\end{figure*}

We consider a spherical grid extending from $\sim$1 to 50~$R_{*}$ with a maximum base resolution of 0.025~$R_*$. To properly capture the eruption and propagation of the CME, the anchoring active region is enclosed by a $45^{\circ}$ (latitude) $\times$ $80^{\circ}$ (longitude) spherical wedge, with twice the maximum base resolution (Fig.~\ref{fig_1}), and reaching up to 25 $R_{*}$. We trace the propagation of each CME inside the stellar corona domain for one hour of physical time\footnote[5]{Computational requirements and the crossing time of a slow ($\sim$ 500 km s$^{-1}$) solar CME event up to a height $> 2.5$ R$_{\odot}$ served to inform the selection of this time-scale. It is good to note here that it doubles the longest temporal evolution considered in the calibration study of the GL model performed by \citetads{2017ApJ...834..173J}.} (with full 3D MHD snapshots every 5 minutes). 

\section{Results}\label{sec_results}

\subsection{Steady-state configuration}

\noindent Figure \ref{fig_2} shows two different views of the resulting steady-state corona and stellar wind solution. As expected, the global structure is dominated by the large-scale component of the driving magnetic field (c.f. Garraffo~et~al. \citeyearads{2013ApJ...764...32G}, \citeyearads{2016A&A...595A.110G}). A standard two-region configuration is obtained, with a fast, low-density wind coming out from the polar regions (open field), complemented with a high-density, slow wind close to the stellar equator (closed field). The solution shows a high-degree of axial symmetry, albeit with some minor deviations introduced by the various mixed-polarity regions corresponding to the small-scale field (Fig. \ref{fig_1}). 

To perform a quantitative characterization of the solution, we calculate the location of the Alfv\'en surface (AS) of the stellar wind. This surface is defined by the collection of points in the 3D domain at which the wind speed ($u_{\rm sw}$) is equal to the Alfv\'en speed ($v_{\rm A}$) of the plasma, defined as $v_{\rm A} \equiv B\,/\sqrt{4\pi\rho}$, where $\rho$ and $B$ are the local density and magnetic field strength, respectively. 

The AS determines the location at which the stellar wind becomes magnetically decoupled from the star (see \citeads{1967ApJ...148..217W}, Mestel~\citeyearads{1968MNRAS.138..359M}, \citeyearads{1999stma.book.....M}, \citeads{1987MNRAS.226...57M}, \citeads{1988ApJ...333..236K}). For this reason, the AS is used to reliably determine the loss rates of mass ($\dot{M}_{*}$) and angular momentum ($\dot{J}_{*}$) from the MHD solution (c.f. \citeads{2014ApJ...783...55C}, Vidotto et al. \citeyearads{2014MNRAS.438.1162V}, \citeyearads{2015MNRAS.449.4117V}, Garraffo~et~al. \citeyear{2015ApJ...813...40G}, \citeyearads{2016A&A...595A.110G}). Our simulations indicate a maximum AS extent of $30~R_{*}$ close to the polar regions, with an average size of $25 R_{*}$. In contrast, an equivalent solar wind simulation using the unmodified CR 2107 magnetic field map yields a mean AS radius of $15$ $R_{*}$. 

\begin{deluxetable*}{cCCCCrrCCCC}[t!]
\tablecaption{Information for the different runs of the GL flux rope CME model.\label{tab_2}}
\tablecolumns{12}
\tablenum{2}
\tablewidth{0pt}
\tablehead{
\colhead{Run \#} & \colhead{$r_0$} & \colhead{$a_1$} & \colhead{$\Phi_{\rm p}$} & \colhead{$E_{\rm B}^{\rm FR}$} & \colhead{$R^{\rm CME}$} & \colhead{$u_{\rm r}^{\rm CME}$} & \colhead{$M^{\rm CME}$} & \colhead{$K^{\rm CME}$} & \colhead{$E^{\rm FL}$} & \colhead{Status}\\
\colhead{} & \colhead{[$R_{*}$]} & \colhead{[G $R_*^{-2}$]} & \colhead{[Mx]} & \colhead{[erg]} & \colhead{[$R_{*}$]} & \colhead{[km s$^{-1}$]} & \colhead{[g]} & \colhead{[erg]} & \colhead{[erg]} & \colhead{}
}
\startdata
01 & 0.4 & 2.2 & 1.13 \times 10^{21}  & 2.45 \times 10^{33} & 2.06   & 176  & 1.59 \times 10^{15} & 2.46 \times 10^{29} & 1.57 \times 10^{30} & C\\
02 & 0.4 & 4.0 & 2.02 \times 10^{21}  & 2.47 \times 10^{33} & 2.08   & 180  & 2.40 \times 10^{15} & 3.88 \times 10^{29} & 3.41 \times 10^{30} & C\\
03 & 0.5 & 3.5 & 4.31 \times 10^{21}  & 2.49 \times 10^{33} & 2.11   & 185  & 4.11 \times 10^{15} & 7.02 \times 10^{29} & 9.58 \times 10^{30} & C\\
04 & 0.8 & 1.6 & 1.29 \times 10^{22}  & 3.14 \times 10^{33} & 2.90   & 338  & 1.04 \times 10^{16} & 5.94 \times 10^{30} & 4.27 \times 10^{31} & C\\
05 & 0.8 & 2.4 & 1.94 \times 10^{22}  & 6.80 \times 10^{33} & 3.46   & 446  & 3.61 \times 10^{16} & 3.59 \times 10^{31} & 7.42 \times 10^{31} & C\\
06 & 1.0 & 2.0 & 3.94 \times 10^{22}  & 2.03 \times 10^{34} & 4.99   & 741  & 8.01 \times 10^{16} & 2.20 \times 10^{32} & 1.95 \times 10^{32} & C\\
07 & 1.2 & 2.4 & 9.80 \times 10^{22}  & 8.62 \times 10^{34} & 5.75   & 889  & 3.12 \times 10^{17} & 1.23 \times 10^{33} & 6.76 \times 10^{32} & E\\
08 & 1.2 & 3.0 & 1.44 \times 10^{23}  & 1.10 \times 10^{35} & 10.26 & 1761 & 4.15 \times 10^{17} & 6.44 \times 10^{33} & 1.14 \times 10^{33} & E\\
09 & 1.6 & 1.6 & 2.06 \times 10^{23}  & 2.50 \times 10^{35} & 11.90 & 2077 & 5.81 \times 10^{17} & 1.25 \times 10^{34} & 1.87 \times 10^{33} & E\\
10 & 1.6 & 2.4 & 3.10 \times 10^{23}  & 4.19 \times 10^{35} & 17.09 & 3080 & 9.27 \times 10^{17} & 4.40 \times 10^{34} & 3.24 \times 10^{33} & E\\
11 & 1.5 & 4.0 & 3.99 \times 10^{23}  & 4.85 \times 10^{35} & 17.60 & 3179 & 1.15 \times 10^{18} & 5.82 \times 10^{34} & 4.57 \times 10^{33} & E\\
12 & 2.0 & 2.4 & 7.56 \times 10^{23} & 1.64 \times 10^{36} & 18.72 & 3395 & 2.05 \times 10^{18} & 1.18 \times 10^{35} & 1.10 \times 10^{34} & E\\
\enddata
\tablecomments{Columns $1-5$ correspond to the Run number, the size ($r_0$), magnetic strength ($a_1$), poloidal flux ($\Phi_{\rm p}$), and associated magnetic energy ($E_{\rm B}^{\rm FR}$), of the erupting GL flux rope. Columns $6-9$ contain derived CME quantities, including maximum values for traveled distance ($R^{\rm CME}$), radial speed ($u_{\rm r}^{\rm CME}$), mass ($M^{\rm CME}$), and kinetic energy ($K^{\rm CME}$), after one hour of evolution. Column $10$ lists the total flare energy ($E^{\rm FL}$) of each event, as estimated in Sect.~\ref{sec_discussion}. The ``Status''  column indicates whether the CME was \textit{confined} (C) or managed to \textit{escape} (E) the large-scale magnetic field. \clearpage}
\end{deluxetable*}

In addition, the solution indicates a maximum radial wind speed of $835$~km~s$^{-1}$, with an associated mass loss rate of $\dot{M}_{*} \simeq 1.59 \times 10^{13}$~g~s$^{-1}$ (which is equivalent to $\sim$\,$2.5 \times\,10^{-13}$~$M_{\odot}$~yr$^{-1} \simeq$\footnote[6]{Assuming $\dot{M}_{\odot} = 2 \times 10^{-14}$~$M_{\odot}$~yr$^{-1}$.}  $12.5$~$\dot{M}_{\odot}$). Despite the generic field configuration driving the simulation, the resulting $\dot{M}_{*}$ is comparable with the estimates from astrospheric Lyman-$\alpha$ absorption reported by \citeads{2005ApJ...628L.143W}, for stars with commensurate large-scale magnetic field strengths (e.g. $\xi$ Boo A, \citeads{2012A&A...540A.138M}; $\epsilon$ Eri, Jeffers~et~al.~\citeyearads{2014A&A...569A..79J}, \citeyearads{2017MNRAS.471L..96J}). 

\subsection{Confined CMEs}\label{Confined_CMEs}

\noindent As described in Sect.~\ref{sec_models}, our CME simulations consider GL flux rope eruptions with increasing poloidal flux $\Phi_{\rm p}$ (see Table \ref{tab_2}). We start by exploring the extent of $\Phi_{\rm p}$ values employed for solar simulations (from $1.0 \times 10^{21}$~Mx to $\sim$\,$2.2 \times 10^{22}$ Mx; \citeads{2017ApJ...834..173J}), yielding the observed range of CME speeds (between $\sim 750 - 3200$~km~s$^{-1}$). We will find that these flux rope parameters do not produce such CMEs in the presence of the 75~G large-scale field. Instead, these eruptions are arrested and do not escape, and we refer to them as \textit{confined CMEs}.

\begin{figure*}[!t]
\centering
\includegraphics[width=0.331\textwidth]{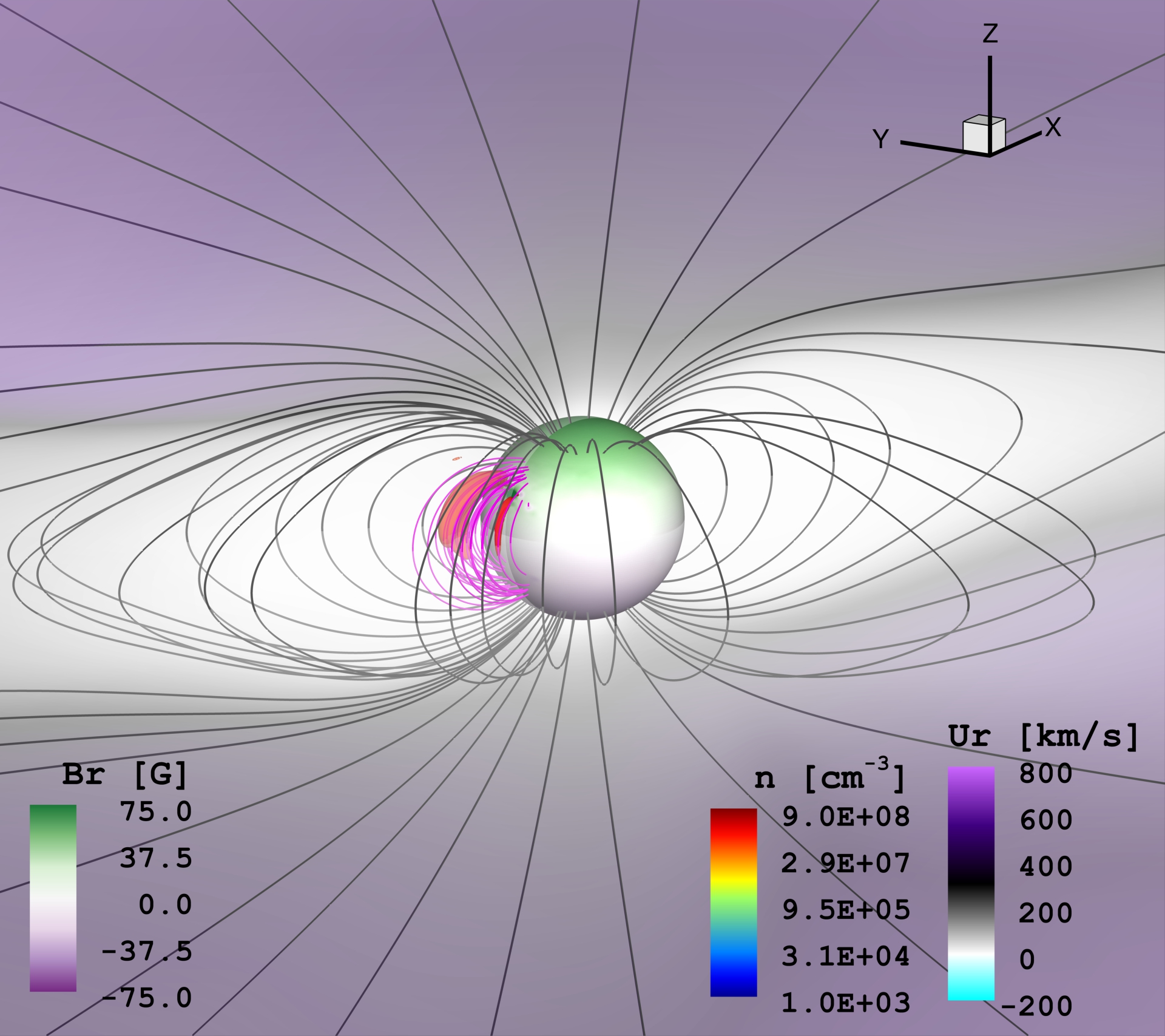}\hspace{1.2pt}\includegraphics[width=0.331\textwidth]{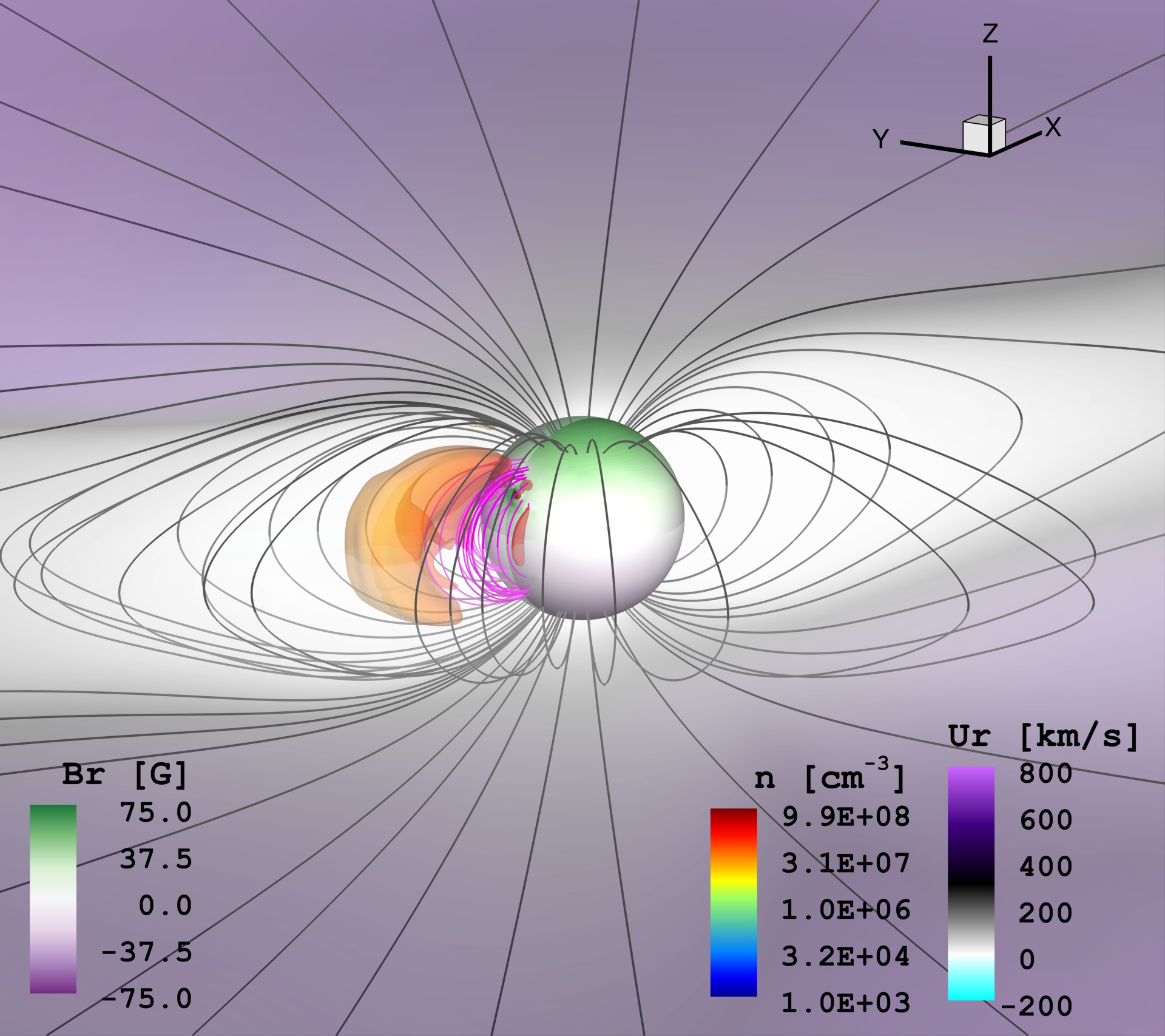}\hspace{1.2pt}\includegraphics[width=0.331\textwidth]{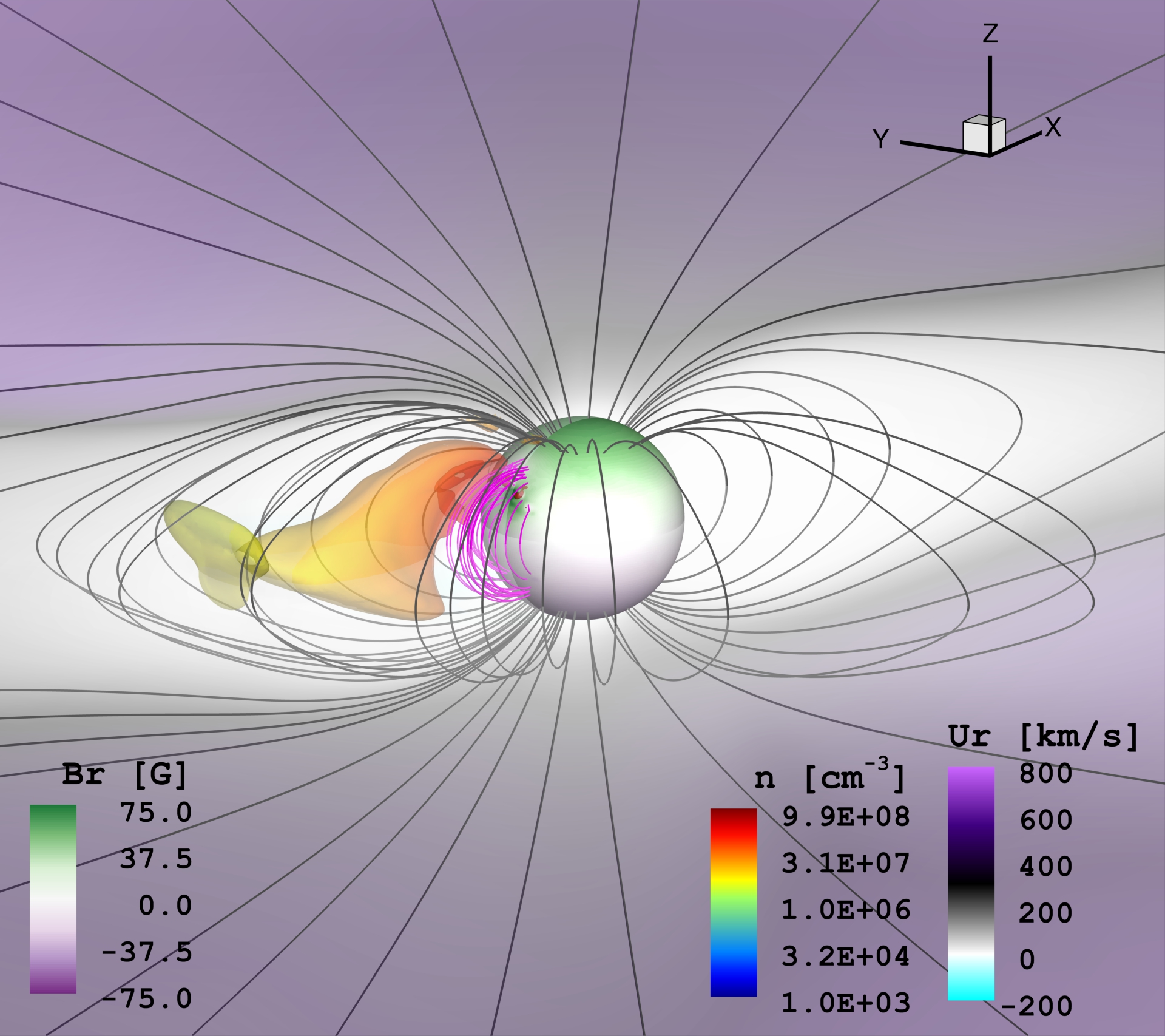}
\includegraphics[width=0.331\textwidth]{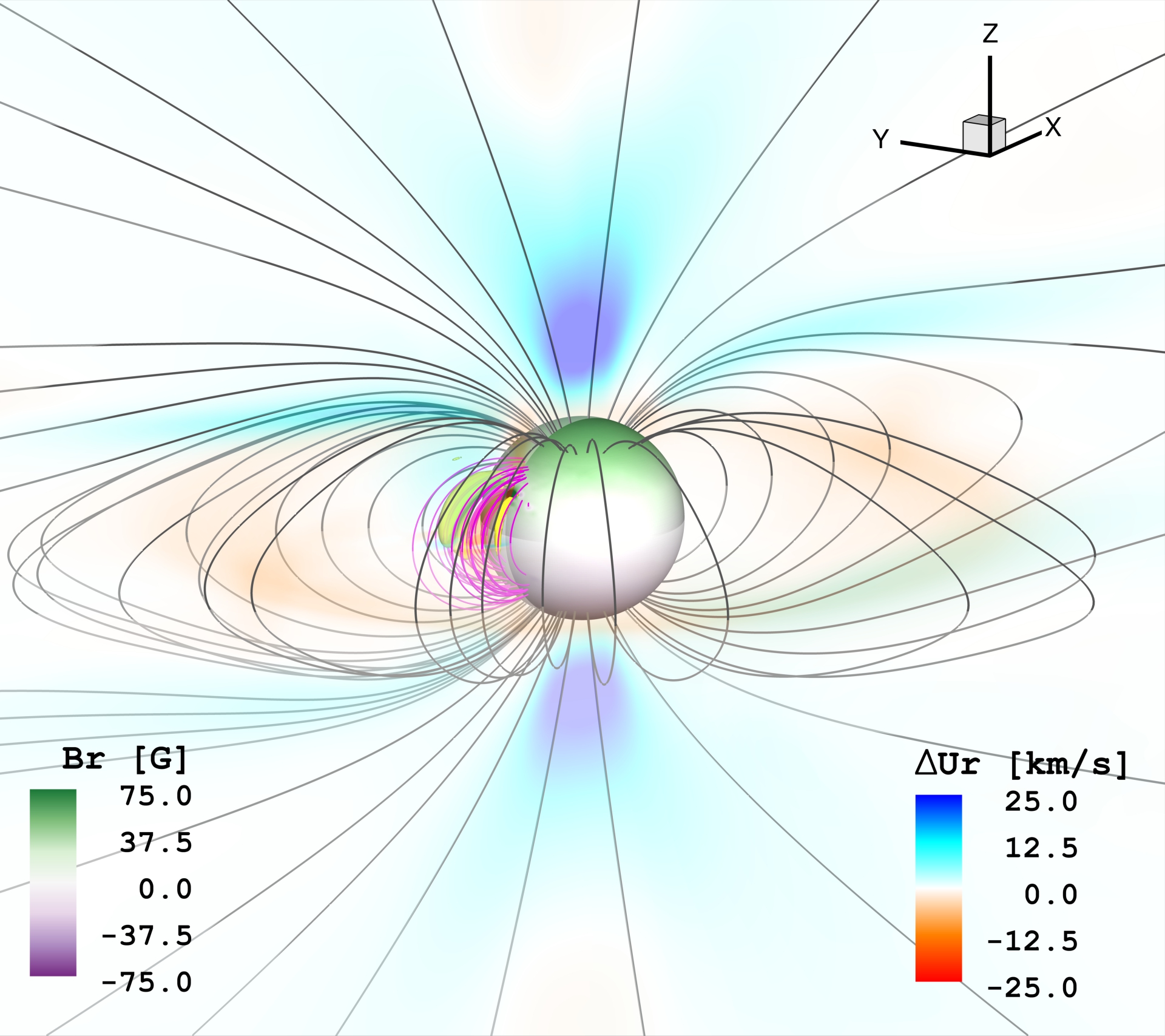}\hspace{1.2pt}\includegraphics[width=0.331\textwidth]{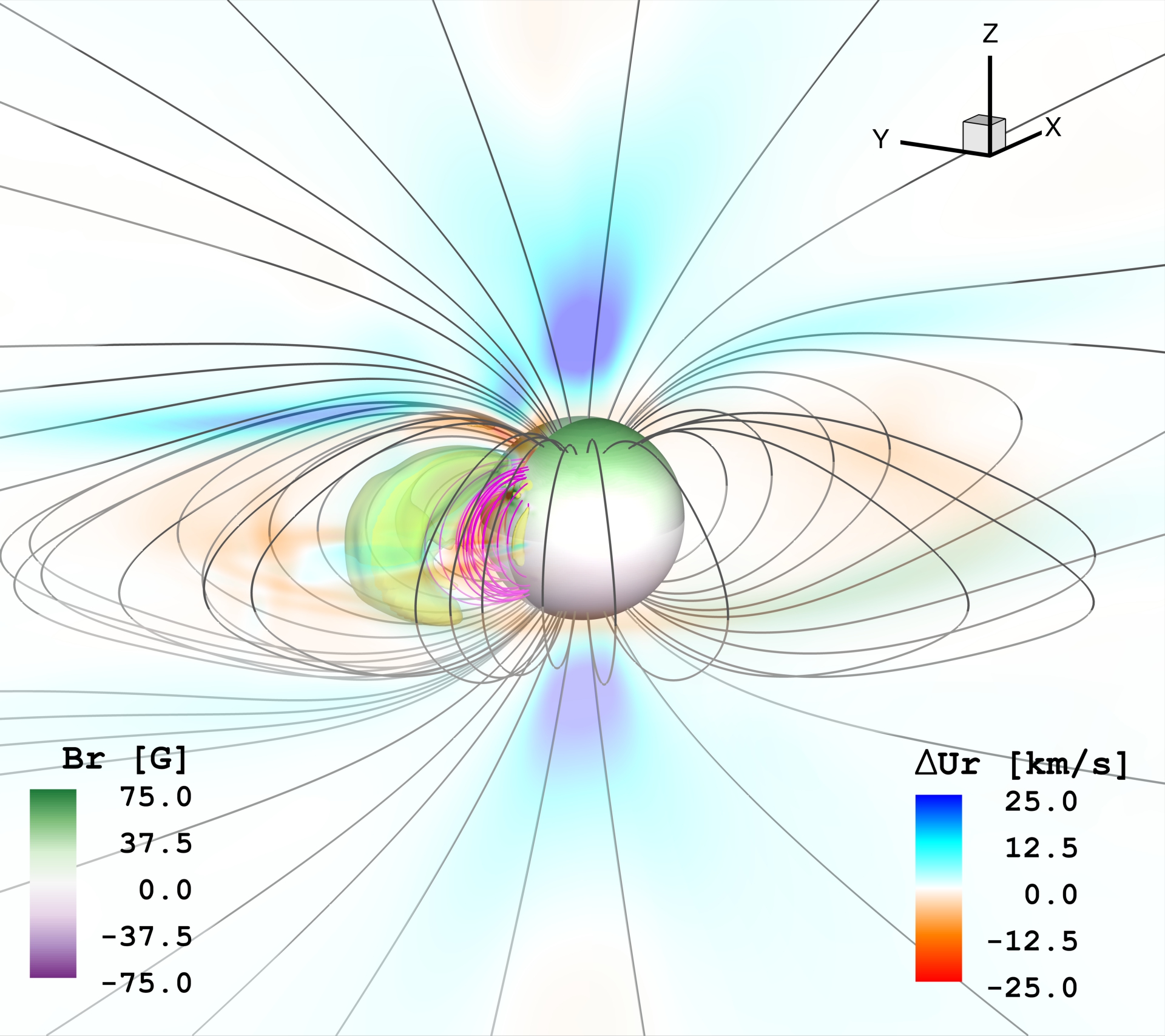}\hspace{1.2pt}\includegraphics[width=0.331\textwidth]{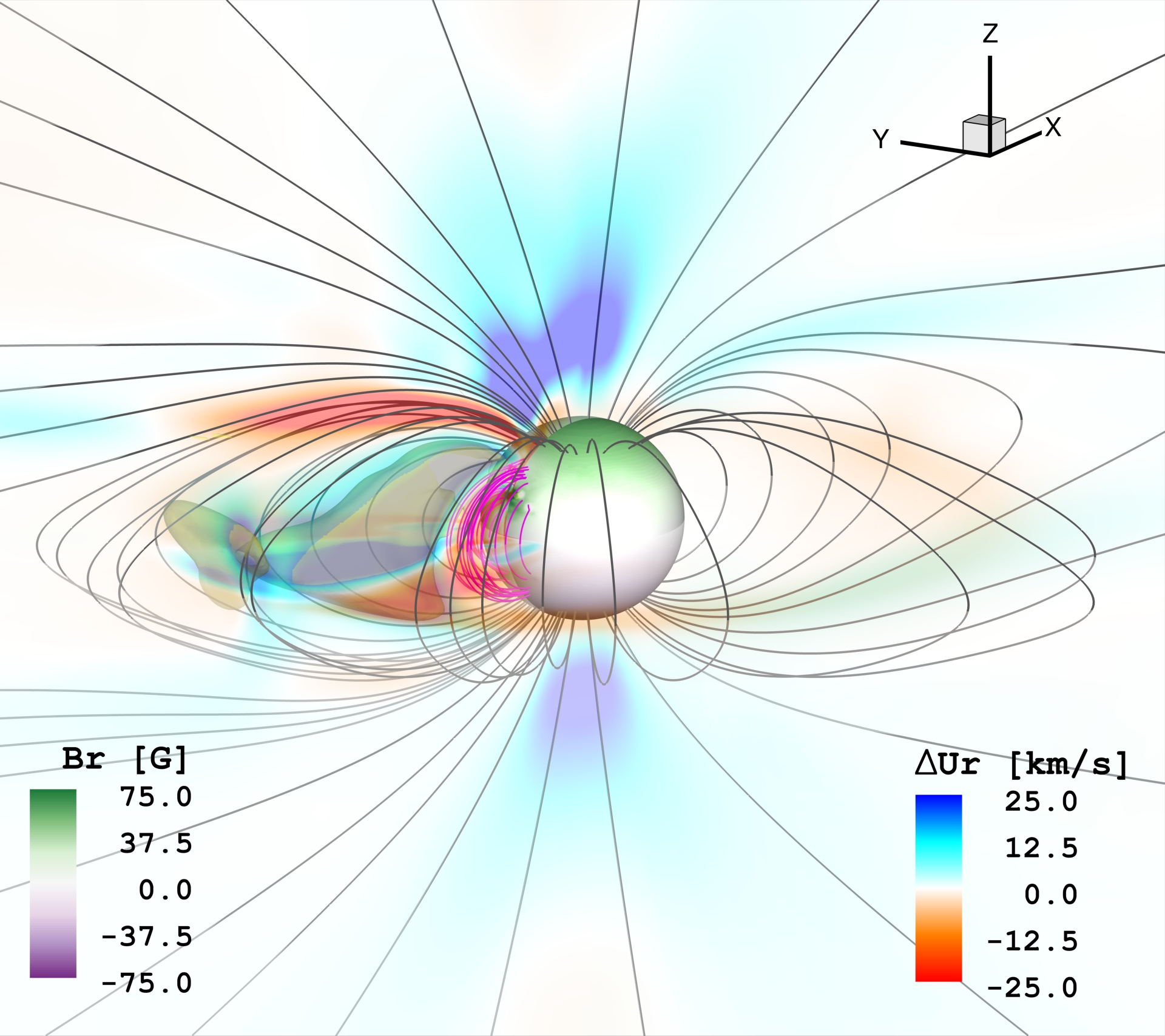}
\caption{Results after one hour of evolution of three different GL flux rope CME simulations. Each column corresponds to a different value of $\Phi_{\rm p}$, associated with the runs 03 (\textit{left}), 05 (\textit{middle}), and 06 (\textit{right}), listed in Table \ref{tab_2}. The perspective shows the eruptive AR towards the north-western limb on the stellar surface (central sphere). A transversal plane crossing the AR, serves to project the distribution of the radial wind speed (\textit{top}), and its variation with respect to the steady-state pre-CME conditions (\textit{bottom}). The identified CME ejecta is shown in all cases, color-coded by plasma density (\textit{top}), and as a translucent yellow shade (\textit{bottom}) for clarity purposes. Grey magnetic field lines are indicative of the large-scale magnetic field, while a selection of field lines seeded around the eruptive AR is shown in magenta. The field of view of all panels is 12~$R_{*}$. }
\label{fig_3}
\end{figure*}

The \textit{left} and \textit{middle} columns of Fig.~\ref{fig_3} show the final state (after one hour of evolution), of two eruptions with $\Phi_{\rm p}$ values at the mid- and high-end of the solar range (Table~\ref{tab_2}, runs 03 and 05). In the \textit{right} panels of Fig.~\ref{fig_3} we present the results of a GL flux rope eruption with roughly twice the maximum poloidal flux value considered in the solar validation study of \citetads{2017ApJ...834..173J}. 

While these GL flux eruptions would have a relatively large impact on the wind for a typical solar magnetic field, this is not the case for the configuration including an additional strong large-scale dipole component. As can be seen from the \textit{top} panels of Fig.~\ref{fig_3}, the ambient stellar wind appears relatively unaltered during the evolution of the CME (density-colored iso-surface in Fig.~\ref{fig_3}, \textit{top}). The latter is identified over the time-dependent simulation by the regions in the 3D domain, which display a simultaneous enhancement of both the local density ($\ge50$\%) and wind speed ($\ge25$\%) with respect to the steady-state conditions. While the eruptions are arrested and do not escape the lower corona, they do not remain static as the simulation evolves. Table~\ref{tab_2} contains the maximum traveled distance, achieved $\sim$\,$30 - 40$ minutes after the onset of the eruptions, and the radial speed in each case. For the remaining simulation time, the confined eruptions largely preserve their shape and location.

In order to visualise any CME-induced changes on the global structure, the \textit{bottom} panels of Fig.~\ref{fig_3} show the difference in radial wind speed ($\Delta u_{\rm r}$) between the pre- and post-CME states (after one hour of evolution). The patches of enhanced wind speed appearing towards the poles are not caused by the CME, but rather by the difference between a steady-state and a time-dependent wind simulation\footnote[7]{This was confirmed by performing the same analysis on a CME-less time-dependent stellar wind simulation.}. On the other hand, it is possible to observe mixed regions of decreasing (red) and increasing (blue) wind speed in the vicinity of the erupting AR, particularly for the strong CMEs presented in Fig.~\ref{fig_3} (\textit{middle} and \textit{right} panels). Both cases display a signature of wind speed reduction, which extends from roughly the maximum height reached by the CME, back to a high-latitude region on the stellar surface ($\sim20^{\circ}$ higher than the initial location of the erupting flux rope, see Table~\ref{tab_1}). In contrast, the largest speed enhancement feature, arising from the change in the confining magnetic field, moves towards the equatorial plane as the simulation evolves. This behaviour was expected since for a given radius, a dipolar large-scale field aligned with the $z-$axis reaches its minimum value around this location. More specifically, the collection of points where $B = 0$ define the current sheet of the system, which for this large-scale configuration is roughly coincident with the equatorial plane, where the outgoing plasma will experience the least amount of resistance by the magnetic field. Similar results have been reported in semi-empirical simulations of solar CMEs (see~Kay~et~al. \citeyearads{2013ApJ...775....5K}, \citeyearads{2015ApJ...805..168K}). 

As was mentioned in Sect. \ref{sec_models}, the initial configuration of the GL model modifies the density profile above the erupting AR. However, no extra mass is added to the flux rope meaning that our simulated CMEs only carry away the ambient coronal plasma. The quantities listed in Table~\ref{tab_2} correspond to the CME mass after one hour of evolution, containing the mass initially perturbed by the eruption above the AR, and the coronal mass swept up during the CME expansion. These have been calculated by performing a numerical integration over the 3D volume defined by the evolving CME (i.e., the above mentioned simultaneous positive gradients in $n$ and $u_{\rm r}$ with respect to the ambient wind solution). It is interesting to note that even without the contribution from the filament, the values derived for all the fully suppressed cases fall within the mass ($10^{12} - 10^{17}$~g) and kinetic energy ($10^{28} - 10^{32}$ ergs) ranges observed in solar CMEs (\citeads{2009EM&P..104..295G}; \citeads{2017ApJ...847...27A}). 

\begin{figure*}[!ht]
\centering
\includegraphics[width=0.495\textwidth]{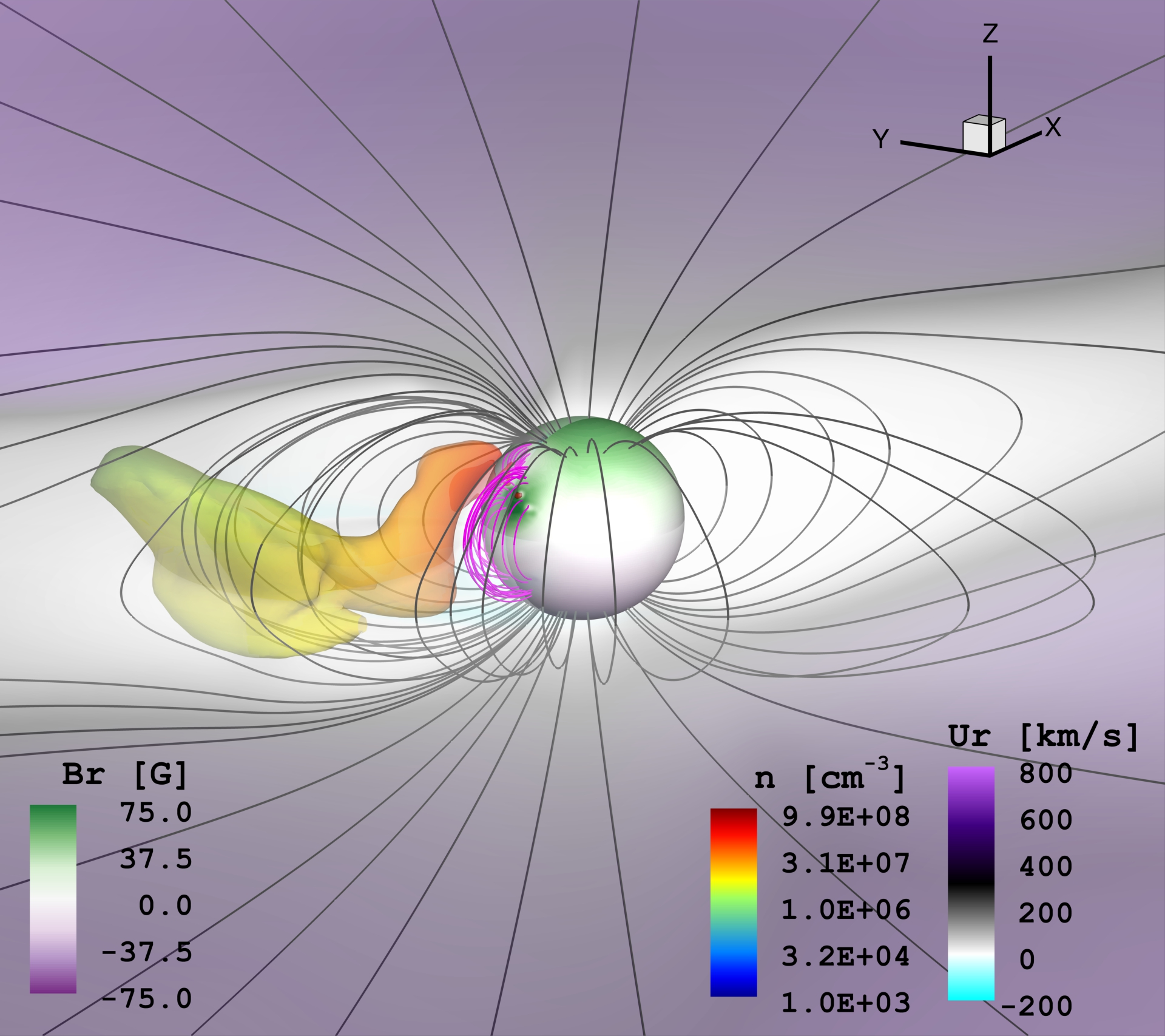}\hspace{1.2pt}\includegraphics[width=0.495\textwidth]{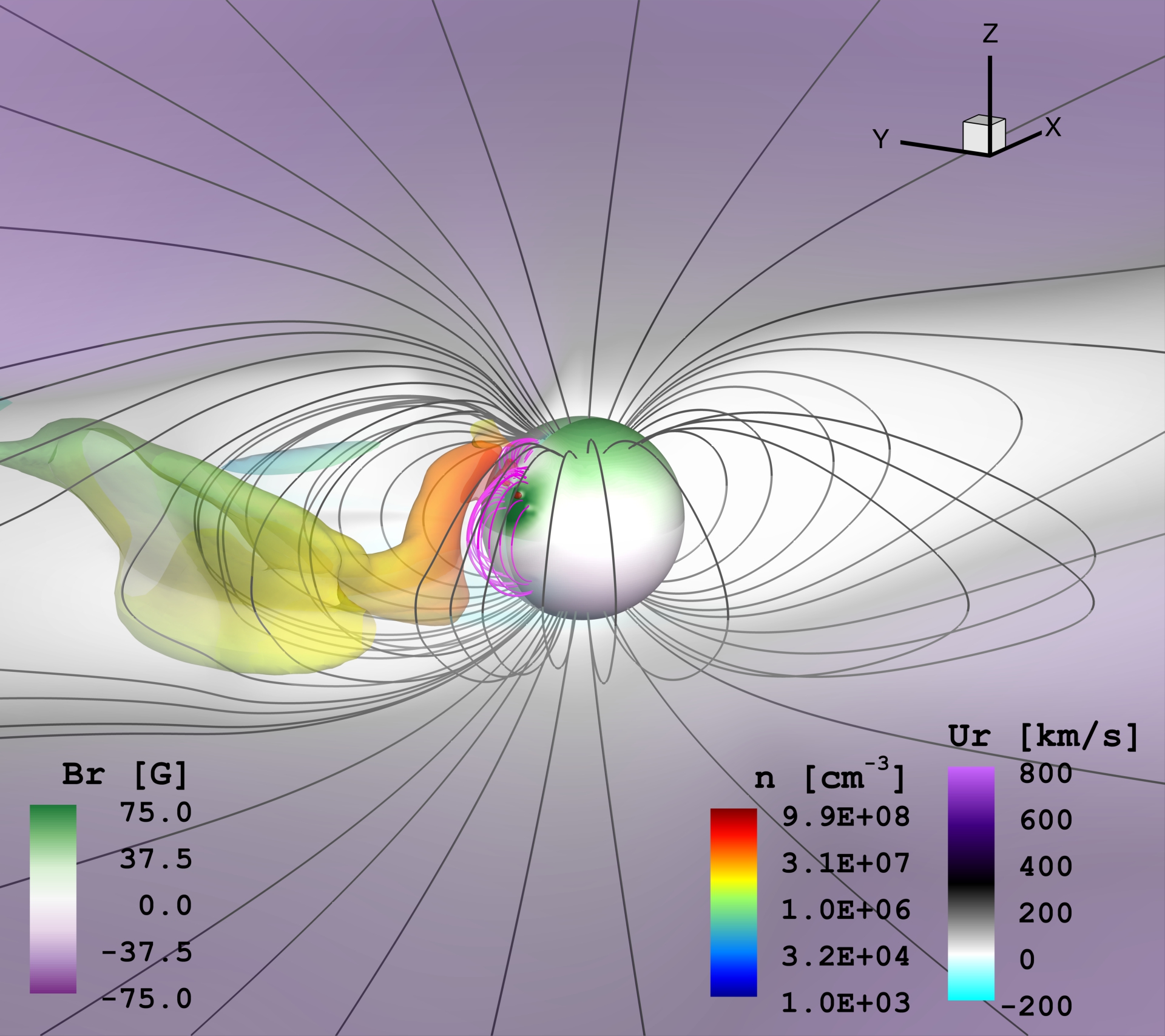}
\includegraphics[width=0.495\textwidth]{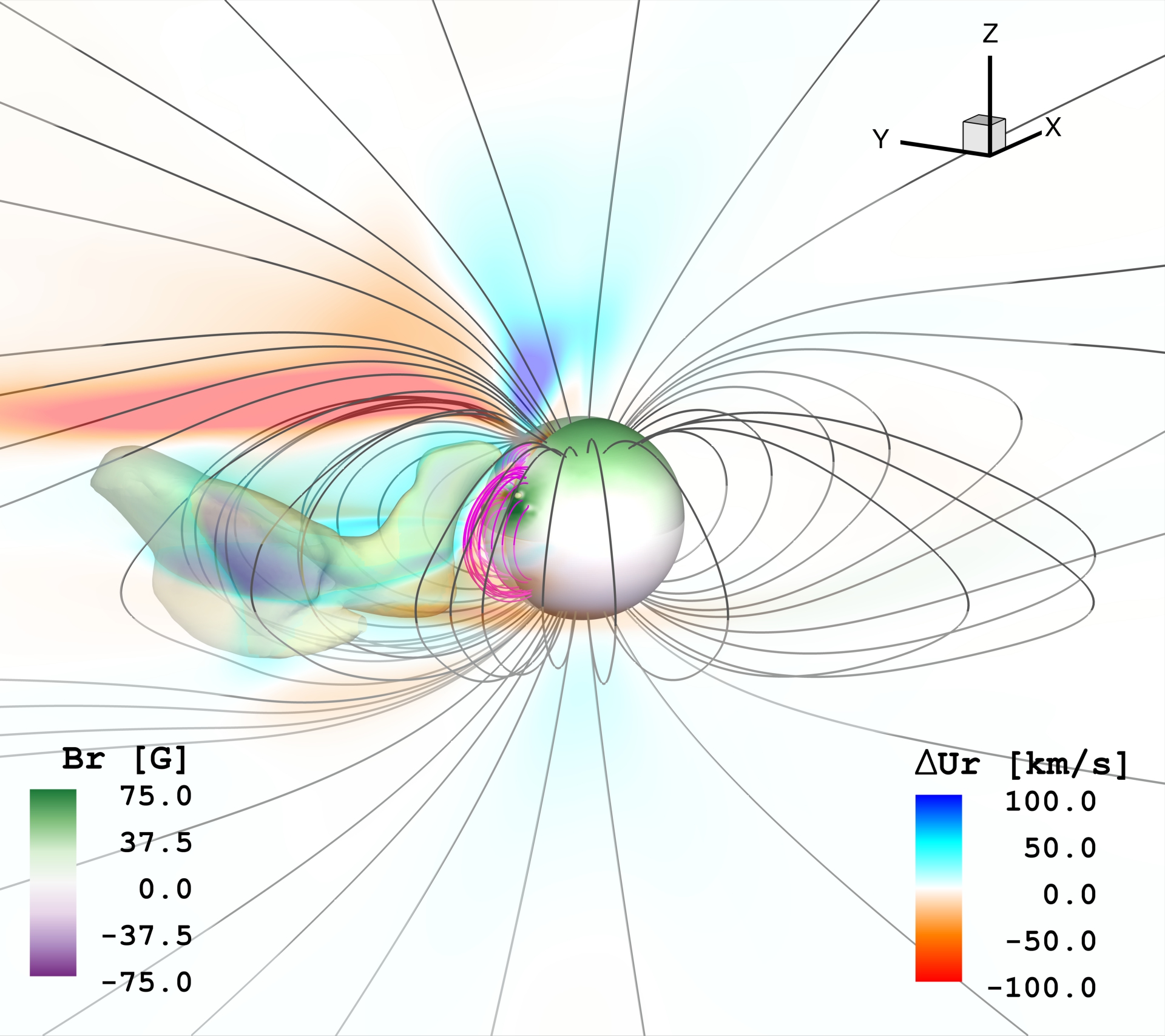}\hspace{1.2pt}\includegraphics[width=0.495\textwidth]{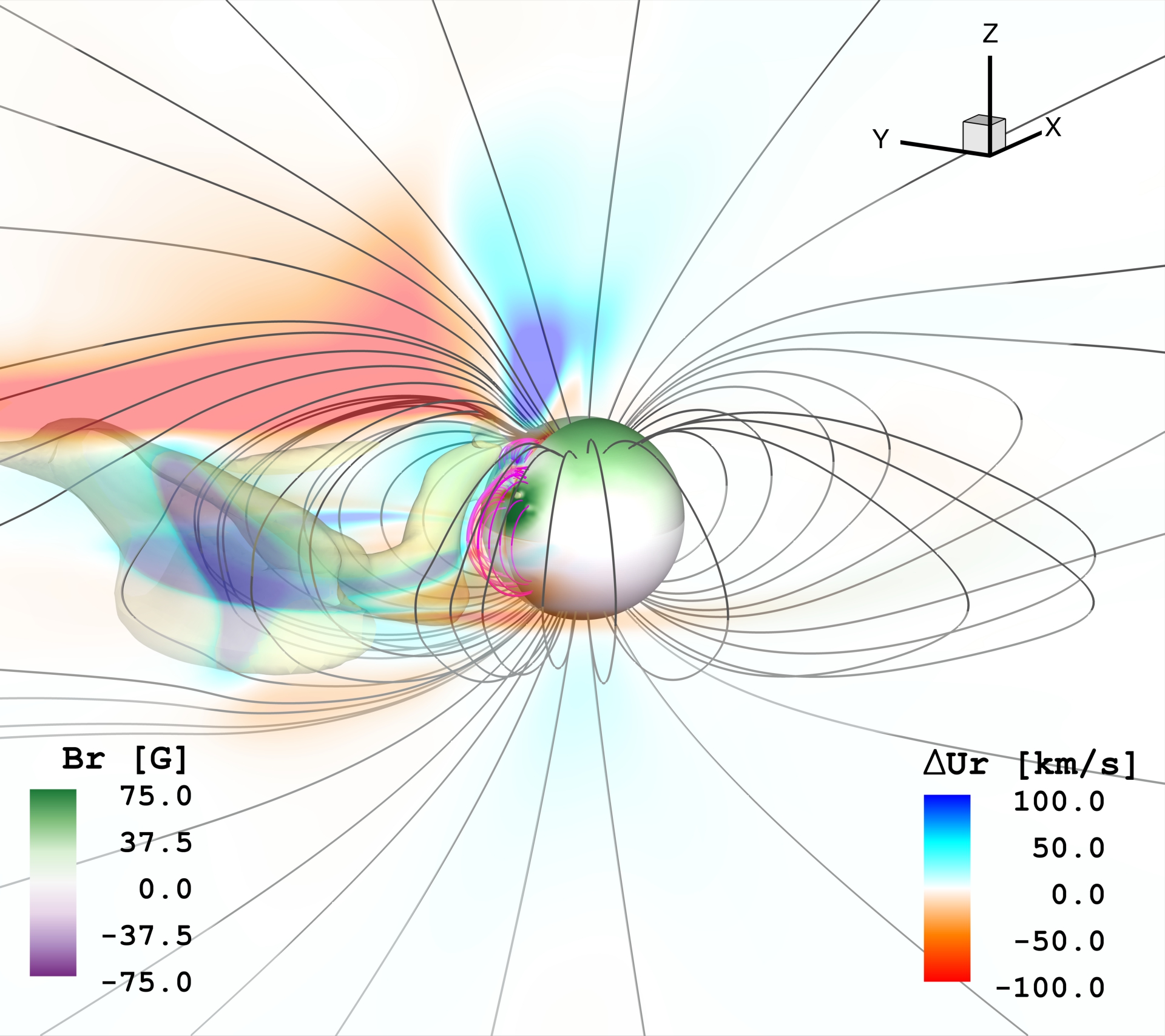}
\caption{Results after one hour of evolution of the GL flux rope CME simulations associated with the runs 07 (\textit{left}) and 08 (\textit{right}) listed in Table \ref{tab_2}. See caption of Fig. \ref{fig_3}.}
\label{fig_4}
\end{figure*}

\begin{figure*}[!ht]
\centering
\includegraphics[width=0.495\textwidth]{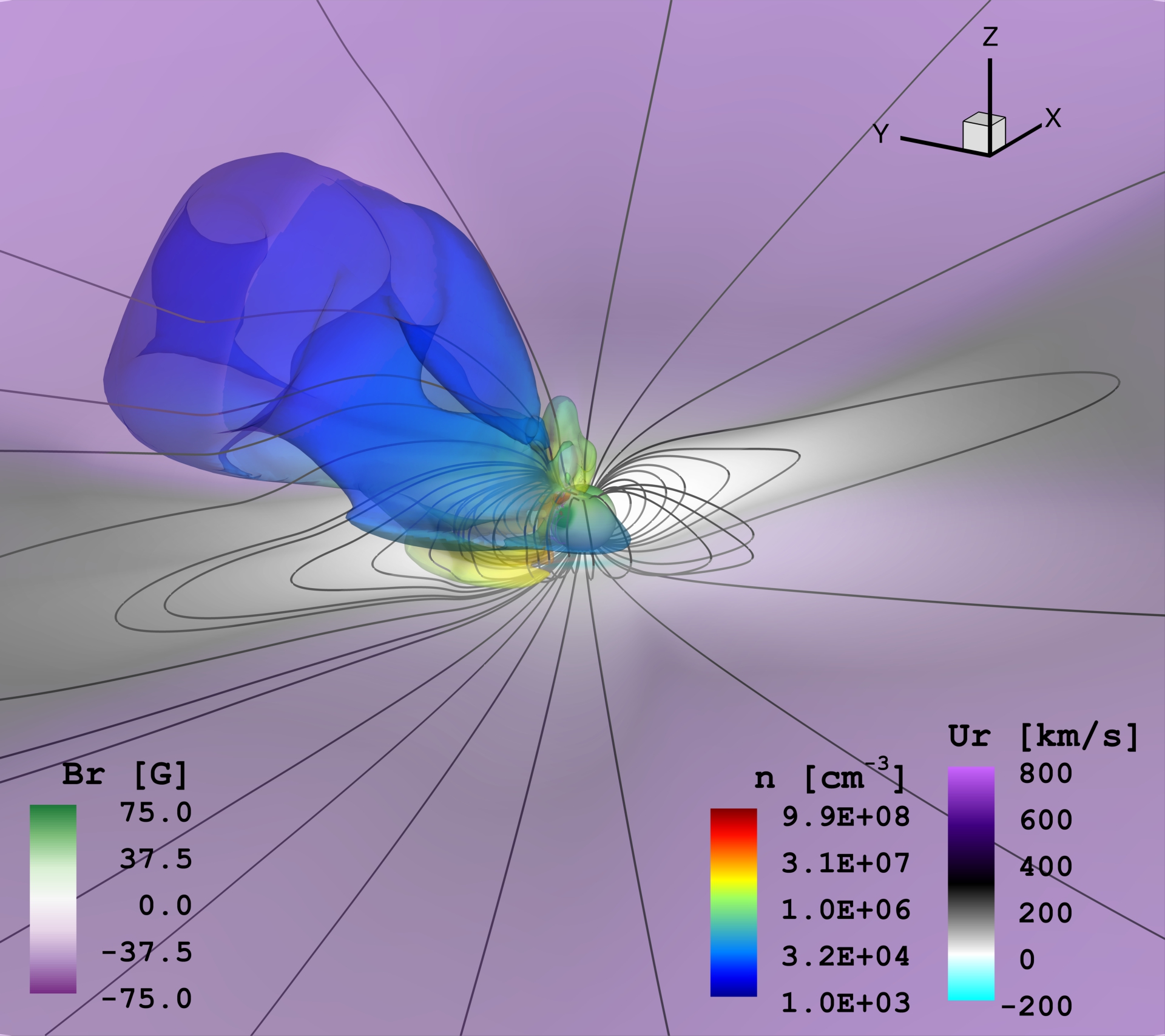}\hspace{1.2pt}\includegraphics[width=0.495\textwidth]{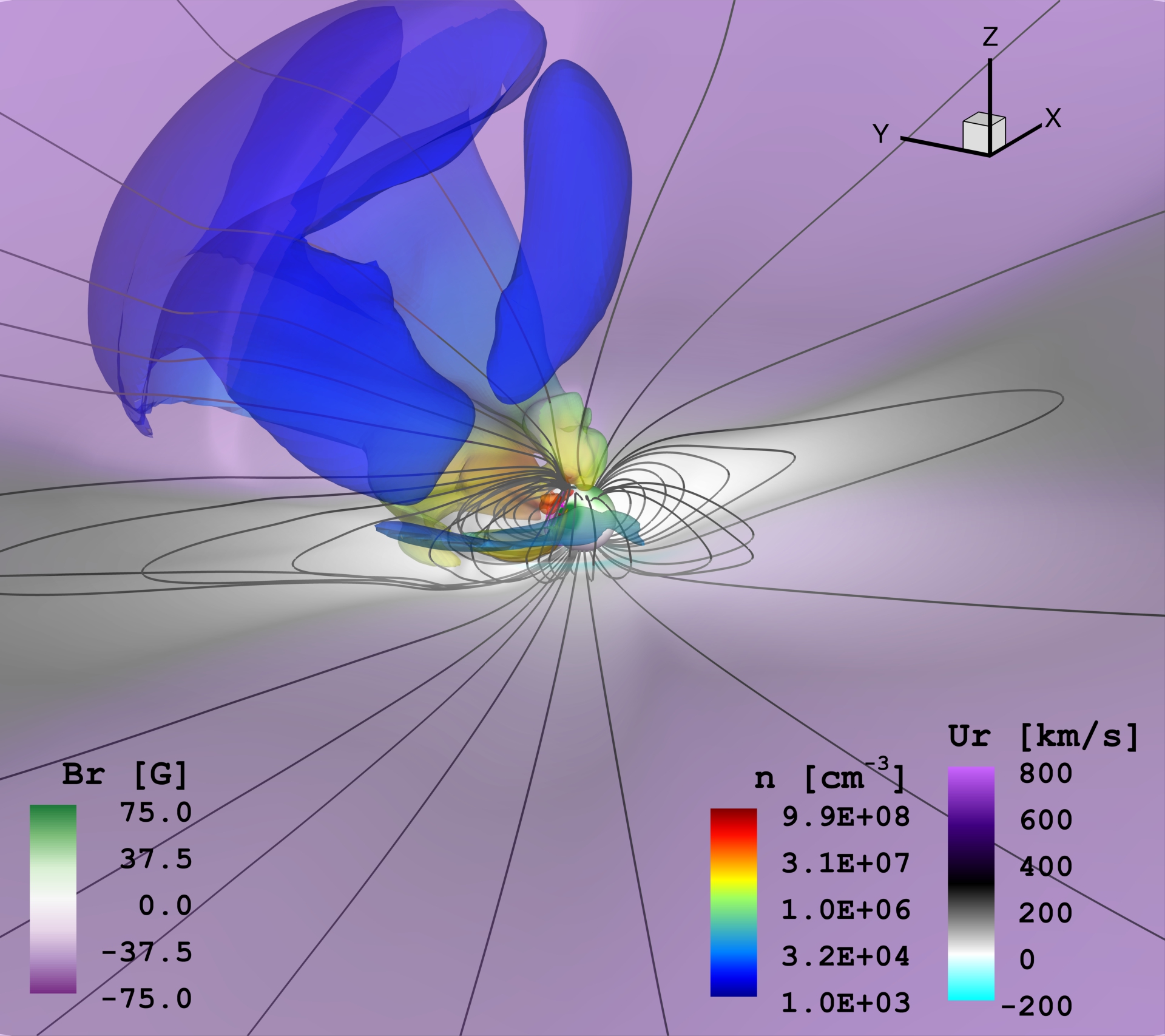}
\includegraphics[width=0.495\textwidth]{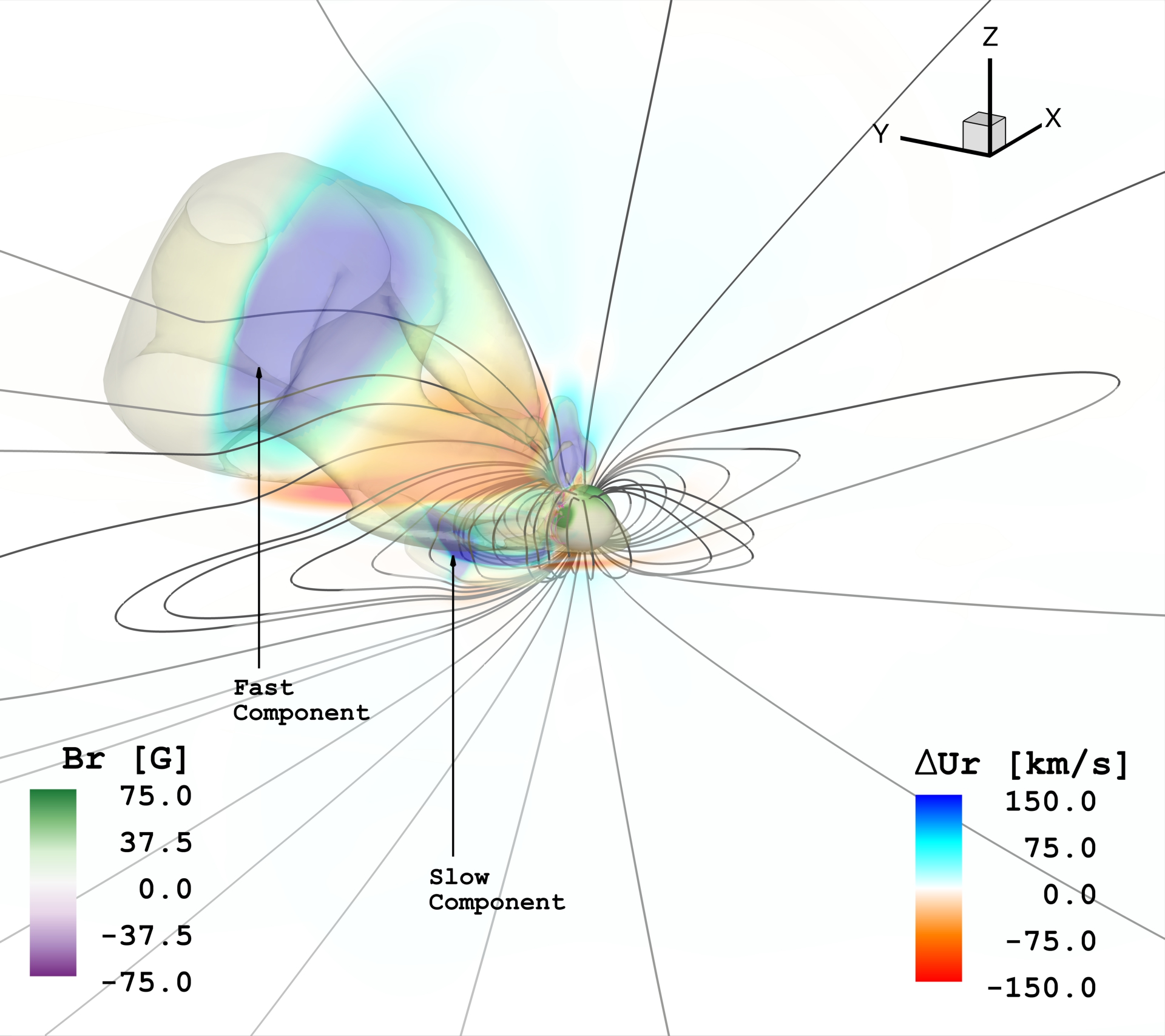}\hspace{1.2pt}\includegraphics[width=0.495\textwidth]{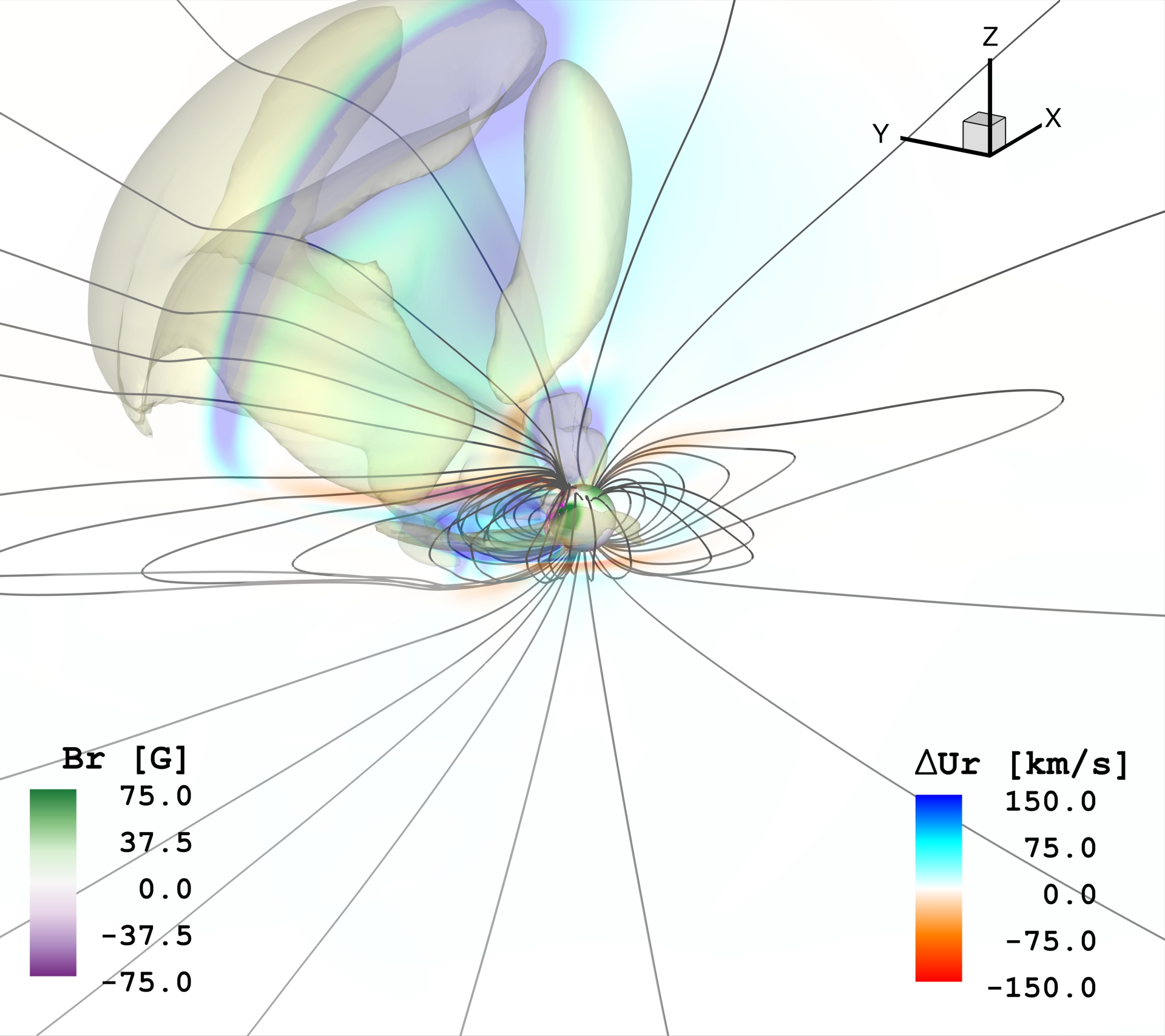}
\caption{Results after one hour of evolution of the GL flux rope CME simulations associated with the runs 11 (\textit{left}) and 12 (\textit{right}) listed in Table \ref{tab_2}. See caption of Fig. \ref{fig_3}. The slow and fast components of the eruption are indicated. The field of view of all panels is 36 $R_{*}$ }
\label{fig_5}
\end{figure*}

Table~\ref{tab_2} also contains the magnetic energy ($E_{\rm B}^{\rm FR}$) associated with each eruption. This quantity has been calculated by performing a numerical integration of the magnetic energy density, $U_{\rm B} = B^{2}/8\pi$, over the volume enclosed by the inner boundary (stellar surface at $r \simeq R_{*}$) and a sphere with radius $R = r_{0} + R_{*}$, and taking the difference between post- and pre-CME states. As expected, $E_{\rm B}^{\rm FR}$ is significantly larger than $K^{\rm CME}$ (up to four orders of magnitude), and increases proportionally to the amount of poloidal flux added by the GL eruption. Our numerical values are fully consistent with observational estimates of the magnetic energy in moderate to large solar flaring events (eruptive and non-eruptive flares with GOES class $\ge$\,M$5.0$; see \citeads{2017ApJ...834...56T}). 

Lastly, for the CME presented in the \textit{right} panel of Fig.~\ref{fig_3}, a small plasmoid can be seen around the apex of the confined loop. This feature could in principle escape over a longer time-scale than the one followed by our simulation. Indeed, its radial velocity with respect to the loop apex is $\sim$\,$320$ km s$^{-1}$, whereas the escape velocity at that location is $\sim$\,$280$ km s$^{-1}$. Nevertheless, as reported in Table~\ref{tab_2}, we have classified this event as a fully suppressed one given that the bulk of the eruption remains magnetically bound to the star. This result indicates that the suppression threshold for our GL flux rope configuration should lie close in parameter space to the values considered for this particular simulation (run~06).

\subsection{Escaping CMEs}\label{Mid_CMEs}

\noindent We proceed by further increasing the amount of poloidal flux of the GL flux rope. Figure~\ref{fig_4} contains the simulation results after one hour of evolution, of two eruptions with roughly five (\textit{left}) and ten (\textit{right}) times the maximum $\Phi_{\rm p}$ value used in solar CME simulations (\citeads{2017ApJ...834..173J}). The visualizations are analogous to the ones presented in Fig.~\ref{fig_3}. In this case however, the effects of the CME on the ambient stellar wind are more pronounced (notice the colour scale change in the \textit{lower} panels of Fig.~\ref{fig_4} with respect to Fig.~\ref{fig_3}). 

Both events show a similar global structure with a CME front carrying a large fraction of the ambient material and escaping the vicinity of the erupting AR towards the current sheet. As the CME expands, the large-scale magnetic field lines above the erupting AR get compressed as they converge toward the stellar surface, and further reducing the velocity of the coronal material at higher latitudes. Unlike the cases presented in Fig.~\ref{fig_3}, this signature is also visible in the opposite hemisphere of the star (below the equatorial plane in the visualizations of Fig.~\ref{fig_4})\footnote[8]{This can be more easily visualized in the movies included as supplementary material.}. Likewise, closer examination of the magnetic field lines anchored at the erupting AR (magenta field lines in Figs.~\ref{fig_3} and \ref{fig_4}) shows a subtle change in connectivity between the escaping (double loop) and the fully suppressed CME cases (single loop). A small portion of the eruption remains close to the stellar surface indicating some partial confinement, as its geometry largely resembles the fully suppressed cases. The radial speeds of these two events are less than 1800~km~s$^{-1}$ and well within the observed range of CME speeds in the Sun. Their associated masses and kinetic energies are also comparable with the maximum values derived from solar observations (i.e.,~$M_{\rm max, \odot}^{\rm CME} \simeq 2.0 \times 10^{17}$~g  and $K_{\rm max, \odot}^{\rm CME} \simeq 4.2 \times 10^{33}$~erg; \citeads{2009EM&P..104..295G}). On the other hand, the associated magnetic energies in these two runs are roughly one order of magnitude larger than the estimated value of a plausible extreme event occurring in the Sun ($\sim$\,$1.5 \times 10^{34}$ erg, \citeads{2017ApJ...834...56T}), and are still more than one order of magnitude larger than the corresponding CME kinetic energy in each case.

\subsection{Monster CMEs}\label{Giant_CMEs}

\noindent We now consider the strongest events in our simulation set (runs 11 and 12 in Table~\ref{tab_2}), which are presented in Fig.~\ref{fig_5}. Apart from a larger field of view ($36$ R$_*$), necessary to capture the rapidly expanding CME, the visualizations are analogous to those presented in Figs.~\ref{fig_3} and \ref{fig_4}. These events are clearly escaping the large-scale magnetic field of the star. However, they show a different spatial structure compared with the relatively weaker eruptions analyzed in the last section. 

After the onset of the eruption, the escaping CME fragments itself into two different parts, which are separated by a region of strong decrement in wind speed (in approximately the same location as in the previous cases). The first one breaks through the large-scale magnetic field and quickly moves to higher latitudes, while the second one slowly drifts towards the equatorial plane, compressing the surrounding field lines in its path. The slightly weaker event associated with run 10 in Table~\ref{tab_2} (not shown in Fig.~\ref{fig_5}), displays a similar behaviour. 

Each CME component moves with a different radial speed, with the first one, at higher latitude, moving approximately 2000~km~s$^{-1}$ faster than the second (see Fig.~\ref{fig_5}). The velocity difference is probably caused by the distribution of the coronal density, which is determined by the dipolar topology of large-scale magnetic field (i.e., low density around the poles and high density in the equatorial regions). Due to this density contrast, roughly $\sim60$\,\% of the CME mass is contained in the slower component, with the remaining $40$\,\% escaping within the fast component (the larger volume compensates to some extent the much lower density). Such a distribution may be different for real stellar CMEs, as our simulations do not include the mass contribution from the dense erupting filament.  

\section{Analysis and Discussion}\label{sec_discussion}

\subsection{Magnetic flux and CME speed}

\noindent With limited information on stellar CMEs, we will focus our discussion in the context of previous solar numerical and observational studies. As mentioned earlier, the events presented in Figs.~\ref{fig_4} (and all the remaining cases with higher $\Phi_{\rm p}$ values; see Table \ref{tab_2}), are outside the parameter space used for CME simulations on the Sun. Nevertheless, as a first order approximation we can estimate the potency of these two eruptions for a typical solar magnetic field/wind configuration, by extrapolating the relation connecting $\Phi_{\rm p}$ and $u_{\rm r}^{\rm CME}$ used in the validation of the GL model (\citeads{2017ApJ...834..173J}). This relation, for the CMEs presented in the \textit{left} and \textit{right} panels of Fig.~\ref{fig_4} suggests radial speeds of roughly $10000$~km~s$^{-1}$ and $15000$~km~s$^{-1}$, respectively. However, in the actual coronal environment the properties of the magnetic field and ambient wind have an influence the behavior of escaping CMEs. Such CME velocities are never realized in practice likely owing to drag and retardation in the outer corona (\citeads{2006SoPh..235..345M}). 

The empirical relation employed above was motivated by the findings of \citetads{2007ApJ...659..758Q}, linking the reconnected magnetic flux during a flare ($\varphi_{\rm FL}$) with the observed CME speed. Very recently, \citetads{2017arXiv171204701T} report a similar trend for a sample of $19$ eruptive solar flares, where $u_{\rm r}^{\rm CME} \simeq (720)\cdot(\varphi_{\rm FL}/10^{22}) + 600$, with $u_{\rm r}^{\rm CME}$ in km s$^{-1}$ and $\varphi_{\rm FL}$ in Mx. Interestingly, this relation closely matches the $u_{\rm r}^{\rm CME}(\Phi_{\rm p})$ dependency obtained by \citetads{2017ApJ...834..173J} with their MHD simulations, suggesting an equivalence between the numerical ($\Phi_{\rm p}$) and observational ($\varphi_{\rm FL}$) fluxes. This presents a way to connect the properties of a simulated CME, driven with a certain $\Phi_{\rm p}$ value, with the expected characteristics of a flare containing the same amount of $\varphi_{\rm FL}$. Following this idea we can invert the scaling relation reported by \citetads{2017arXiv171204701T}, derived from a larger sample (51) of confined and eruptive solar events, to connect $\varphi_{\rm FL} \equiv \Phi_{\rm p}$ with the GOES peak soft X-ray flux of the flare ($F_{\rm SXR}^{\rm FL}$):

\begin{equation}
\log(F_{\rm SXR}^{\rm FL}) = \dfrac{\log(\Phi_{\rm p}) - b}{a}\mbox{\,,} 
\end{equation}

\noindent where $b = 0.580 \pm 0.034$ and $a = 24.21 \pm 0.22$. 

\begin{figure}[!t]
\centering
\includegraphics[trim=0.15cm 0.1cm 0.1cm 0.1cm, clip=true,width=0.49\textwidth]{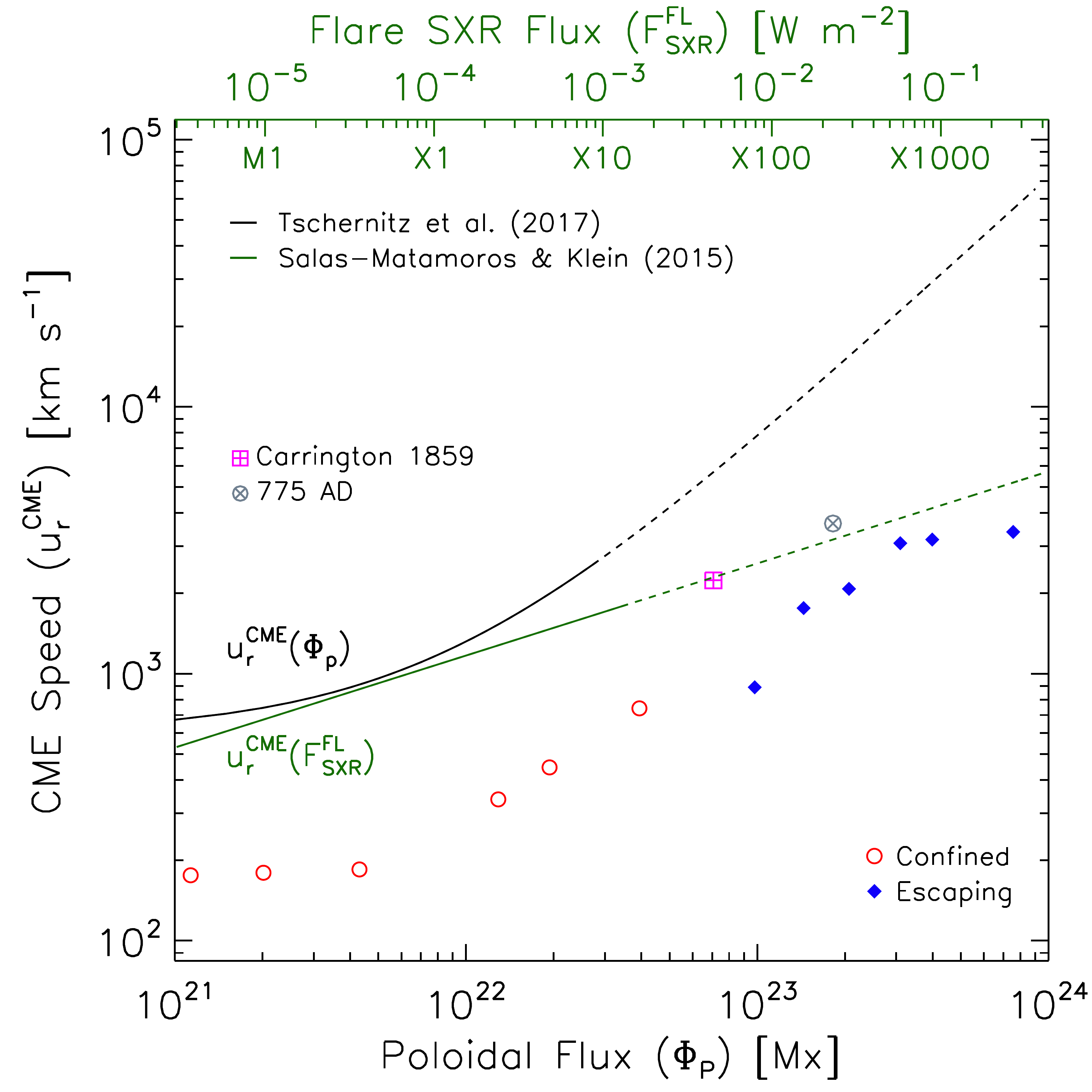}
\caption{Radial speed of our simulated CMEs ($u_{\rm r}^{\rm CME}$) as a function of $\Phi_{\rm p}$. The secondary $x-$axis, showing the flare SXR flux ($F_{\rm SXR}^{\rm FL}$) and corresponding GOES class, has been constructed assuming the equivalence of $\Phi_{\rm p}$ with the reconnection flux $\varphi_{\rm FL}$ (see text for more details), and using the scaling relation reported by \citetads{2017arXiv171204701T}. The black and green curves correspond to the observed behaviour of $u_{\rm r}^{\rm CME}$ with respect to $\Phi_{\rm p}$ (\citeads{2017arXiv171204701T}), and $F_{\rm SXR}^{\rm FL}$ (\citeads{2015SoPh..290.1337S}), respectively. The extrapolated region of each relation is indicated by dashed lines. Symbols denote the CME status in the simulations.}
\label{fig_6}
\end{figure}

Using this procedure, we present in Fig.~\ref{fig_6} the behaviour of $u_{\rm r}^{\rm CME}$ as a function of $\Phi_{\rm p}$ in our simulations, with the corresponding $F_{\rm SXR}^{\rm FL}$ values indicated by an additional $x-$axis. The full extent of the parameter space covered by our simulations is much clearer now, with eruptions of equivalent SXR fluxes ranging between C3.5 and X3000.0 in the GOES classification (nearly five orders of magnitude of difference). This secondary scale allows a direct comparison with the estimated parameters of two historical extreme eruptions in the Sun: the Carrington 1859 flare ($\sim$X45.0, $M^{\rm CME} \sim 8.0 \times 10^{16}$~g, $K^{\rm CME} \sim 2.0 \times 10^{33}$~erg; \citeads{2013JSWSC...3A..31C}) and the 775 AD event\footnote[9]{As indicated by  \citetads{2014ApJ...781...32C}, these are the required parameters in a solar origin explanation for the $^{14}$C concentration increase in tree rings reported by \citetads{2012Natur.486..240M}. Whether or not this space weather event occurred is still under debate.} ($\sim$X230.0, $M^{\rm CME} \sim 7.5 \times 10^{17}$~g, $K^{\rm CME} \sim 5.0 \times 10^{34}$~erg; \citeads{2012Natur.491E...1M}, \citeads{2014ApJ...781...32C})\footnote[10]{Conversely, this also indicates associated poloidal fluxes of $7.0~\times~10^{22}$~Mx for the Carrington 1859 flare, and $\sim$$1.8 \times 10^{23}$ Mx for the 775 AD event.}. Likewise, we have also included the observational results from \citetads{2015SoPh..290.1337S} and \citetads{2017arXiv171204701T}, linking $u_{\rm r}^{\rm CME}$ with $F_{\rm SXR}^{\rm FL}$ and $\varphi_{\rm FL}$,\footnote[11]{We have preserved here the notation used by \citetads{2017arXiv171204701T}. The relation shown in Fig.~\ref{fig_6} uses $\varphi_{\rm FL} = \Phi_{\rm p}$, as was assumed beforehand.} respectively. 

Various aspects of Fig.~\ref{fig_6} are noteworthy. First of all, it is clear that our simulated CMEs are slower than expected from solar events with equivalent magnetic and/or flaring properties. The same is true for events outside the solar parameter space. For instance, the speeds of our strongest events are comparable with the fastest solar CMEs (see Table~\ref{tab_2}). Still, the poloidal fluxes driving these eruptions are more than one order of magnitude larger than what is needed to power a solar CME with the same speed (\citeads{2017ApJ...834..173J}). We attribute this to the fact that all our cases have, to some extent, been suppressed by the large-scale magnetic field. Its good to note here that our methodology for determining the CME speeds is relatively similar to the linear speed determination of the LASCO CME catalog (based on coronograph data between $1.5 - 30$~R$_{\odot}$). 

On the other hand, while the observational relationships are roughly consistent within their ranges of derivation (solid lines in Fig.~\ref{fig_6}), the location of the Carrington and the 775~AD events suggest that the extrapolation of $u_{\rm r}^{\rm CME}(F_{\rm SXR}^{\rm FL})$ is more robust than that of $u_{\rm r}^{\rm CME}(\Phi_{\rm p})$. Under this consideration, the difference between the expected and the simulated CME speeds is larger for the confined events than in the escaping counterparts. As discussed later, this will have important consequences in terms of the kinetic energy carried away by the CMEs. Furthermore, it is interesting to see that the radial speeds of our three strongest cases differ by less than ten percent, despite an increment greater than a factor of 2.4 in $\Phi_{\rm p}$ (almost an order of magnitude in $F_{\rm SXR}^{\rm FL}$). Given the appropriate escaping conditions, this is indicative of a very important (or even dominant) role played by the stellar wind topology in determining the final properties of the eruption, as all these three events displayed the CME fragmentation previously described in Sect.~\ref{Giant_CMEs}. 

Finally, if we consider that the poloidal flux of an eruption is contained within the average area of an AR in the Sun (i.e.~$\sim 4.5 \times 10^{19}$~cm$^2$)\footnote[12]{This corresponds to 1500 millionths of the visible solar hemisphere (\textmu Hem)}, the weaker escaping events in Fig.~\ref{fig_7} (cases 07 and 08 shown in Fig.~\ref{fig_4}) would require flux ropes with field strengths between $2.2$~kG and $3.2$~kG, respectively. These values are up to five times larger than the estimates presented by \citetads{2015ApJ...804L..28S}, of the field strength along the main polarity inversion line (PIL; where the flux ropes reside) of three major solar ARs. The required flux rope field strengths are instead comparable with the PIL estimates for the largest ever reported solar AR\footnote[13]{Observed on Apr. 3 1947 with an area of 6132 \textmu Hem; See \url{http://solarcyclescience.com/activeregions.html}}  (i.e.,~$\sim1.8 \times 10^{20}$~cm$^2$).  We surmise that the conditions to generate large CME events, while extremely unlikely, are not completely excluded in the case of the Sun. This is consistent with the relative location of the Carrington 1859 and the 775 AD events in Fig.~\ref{fig_7}, whose radial speeds would not have been reduced due to the lack of a significant large-scale confining field in the Sun.  

\subsection{CME masses}

\noindent As discussed earlier, the masses associated with the events presented in Sects.~\ref{Confined_CMEs} and \ref{Mid_CMEs} are within nominal solar ranges. On the contrary, the largest (and fastest) CMEs from Sect.~\ref{Giant_CMEs} display masses that are two orders of magnitude larger than the typical values associated with fast CMEs in the Sun (with $u_{\rm r, \odot}^{\rm CME} > 2500$~km~s$^{-1}$), and roughly ten times above the estimates for the most massive event in the the latest version of the SoHO/LASCO CME catalog\footnote[14]{\url{http://www.lmsal.com/solarsoft/www_getcme_list.html}}. This difference was expected considering the increase in size of the eruption (parameter $r_{0}$ in Table~\ref{tab_2}), and the amount of coronal plasma swept away as the event evolves.

\begin{figure}[t!]
\centering
\includegraphics[trim=0.15cm 0.1cm 0.1cm 0.1cm, clip=true,width=0.49\textwidth]{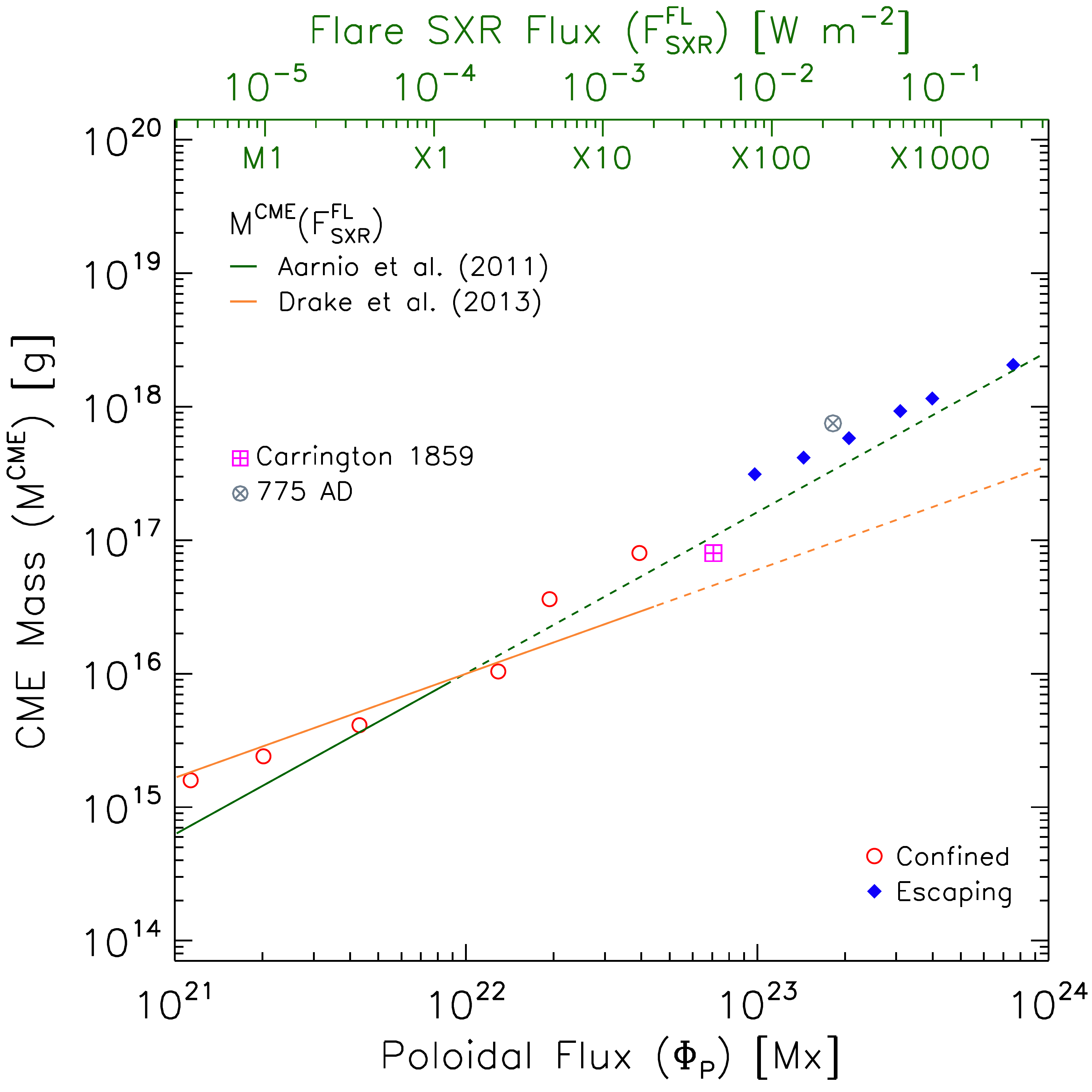}
\caption{Mass of our simulated CMEs ($M^{\rm CME}$) as a function of $\Phi_{\rm p}$. See the caption of Fig.~\ref{fig_6}. As indicated, the solid lines show different relations for $M^{\rm CME}$ with respect to $F_{\rm SXR}^{\rm FL}$ derived from solar observations. Extrapolated ranges for each relation are indicated by dashed lines. Symbols denote the CME status in the simulations.}
\label{fig_7}
\end{figure}

To understand the global behaviour of the resulting CME mass, in Fig.~\ref{fig_7} we present the simulated $M^{\rm CME}$ values as a function of $\Phi_{\rm p}$. The $F_{\rm SXR}^{\rm FL}$ axis values are determined following the same procedure as in Fig.~\ref{fig_6}. 

We include two $M^{\rm CME}(F_{\rm SXR}^{\rm FL})$ relations (and their extrapolations), derived by \citetads{2011SoPh..268..195A} and \citetads{2013ApJ...764..170D}, using similar sets of solar observations. Unlike the radial speeds, the CME masses in our simulations align closely to the observational trends, regardless of their confined or escaping status. In particular, our simulated values seem to follow more closely the \citetads{2011SoPh..268..195A} relation rather than the one from \citetads{2013ApJ...764..170D}. Still, as described in Sects.~\ref{Mid_CMEs} and \ref{Giant_CMEs}, all of our escaping events show some degree of CME fragmentation, where a fraction of the eruption remains confined by the large-scale field. This confined mass can be as high as 50 per cent of the total mass of the eruption, bringing down the masses of the escaping cases in Fig.~\ref{fig_7} towards the scaling from \citetads{2013ApJ...764..170D}. As the filament contribution is neglected in the simulations, our CME masses could be altered depending on a non-trivial response of such structure to the confining field. Nevertheless, if we include the formal uncertainties of these observational scalings, our numerical values are effectively consistent with both relations.

At face value, these results would indicate that the large-scale field has relatively little influence over the CME mass. However, it is important to note here that the CME masses were not included in the calibration of the GL model by \citetads{2017ApJ...834..173J}. Therefore, it is not possible to guarantee that our confined events would have the same CME masses in the absence of the large-scale field (i.e. under normal solar conditions). Likewise, stellar wind simulations have shown that stronger large-scale fields lead to higher mass loss rates (c.f., \citeads{2015ApJ...807L...6G}, \citeads{2018ApJ...856...53P}), and more dense coronae (c.f. Cohen~et~al.~\citeyearads{2014ApJ...783...55C}, \citeyearads{2017ApJ...834...14C}). It is expected then that escaping CMEs under those conditions would sweep up more mass than in a weaker large-scale field case. For these reasons, a more accurate interpretation of Fig.~\ref{fig_7} would consider instead the coronal mass \textit{perturbed} by the erupting flux rope (not necessarily escaping), which should mostly depend on $\Phi_{\rm p}$ and the small-scale field anchoring the eruption, with little influence from the large-scale magnetic field. On the other hand, it is clear that the large-scale field will drastically affect the amount of mass the star can lose via CMEs, with the explicit example studied here of a $75$~G dipolar field being able to fully suppress solar events up to $\sim$X20 in the GOES class. As discussed by \citetads{2016IAUS..320..196D}, the relative importance of this suppression mechanism in the mass loss budget in active stars will strongly depend on whether small or large flares dominate the properties of the corona, with a larger influence on the former case compared to the latter. 

\subsection{CME kinetic energy and suppression threshold}

\begin{figure}[t!]
\centering
\includegraphics[trim=0.15cm 0.1cm 0.1cm 0.1cm, clip=true,width=0.49\textwidth]{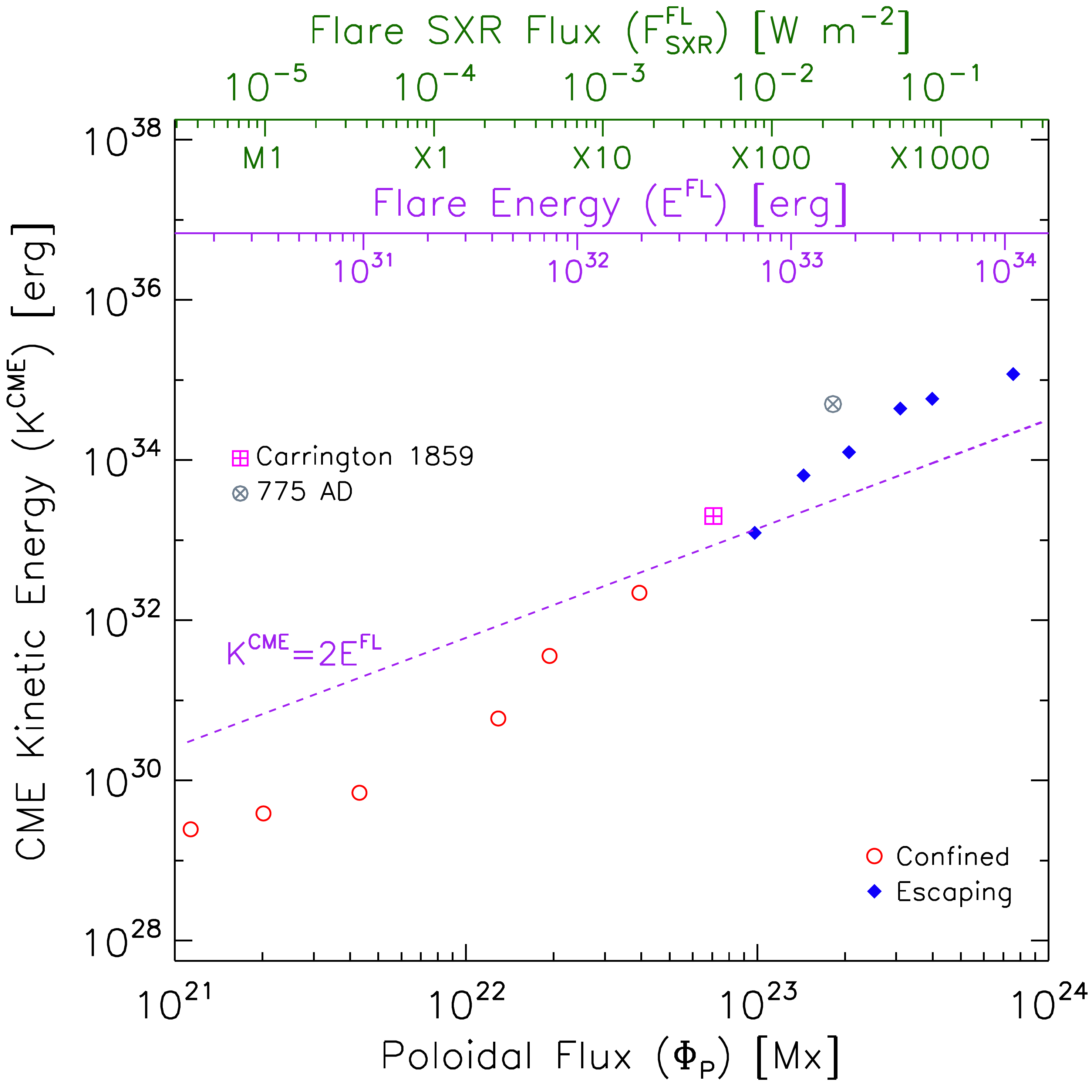}
\caption{Kinetic energy of our simulated CMEs ($K^{\rm CME}$) as a function of $\Phi_{\rm p}$. The secondary $x-$axis showing $F_{\rm SXR}^{\rm FL}$ is constructed as in Fig.~\ref{fig_6}, and serves to compute a third $x-$axis indicating the total flare energy, $E^{\rm FL}$, using the scaling derived by \citetads{2011A&A...530A..84K}. The dashed line shows a 1 to 2 equivalence between CME kinetic and flare energies.}
\label{fig_8}
\end{figure}

\noindent We conclude our discussion by considering the kinetic energy of the CMEs and the suppression threshold established by the large-scale field. Figure~\ref{fig_8} shows $K^{\rm CME}$ in terms of $\Phi_{\rm p}$ in our simulations. The associated $F_{\rm SXR}^{\rm FL}$ values (determined as in Fig.~\ref{fig_6}), are used to compute an additional $x-$axis indicating the total energy of the flare, $E^{\rm FL}$, given by

\begin{equation}\label{eq_eflare}
\log(E^{\rm FL}) = d\cdot\log(F_{\rm SXR}^{\rm FL}) + e\mbox{\,,} 
\end{equation}

\noindent with $d = 0.79 \pm 0.10$ and $e = 34.49 \pm 0.44$. This relation was initially derived in the Sun-as-as-star flaring analysis of \citetads{2011A&A...530A..84K}, and has been shown to be robust even for extreme solar flares (see~\citeads{2016A&A...588A.116W}). Table~\ref{tab_2} contains the derived values of $E^{\rm FL}$ from Eq.~\ref{eq_eflare}, which turn out to be $0.1$ to $1.0$ per cent of the numerically-integrated magnetic energy added by the flux rope ($E_{\rm B}^{\rm FR}$), being the fraction larger for stronger events. Despite the simplicity of this approach, these flare-magnetic energy fractions agree well with typical values observed for the Sun (c.f.~\citeads{2012ApJ...759...71E}). As discussed below, this scaling also allows us to compare the results from our simulations with the characteristic behaviour derived from solar observations. 

By performing a statistical analysis over a database of solar flares and CMEs, \citetads{2013ApJ...764..170D} found that, on average, $K^{\rm CME} \simeq 200\,E_{\rm X}^{\rm FL}$, with $E_{\rm X}$ as the energy of the flare in the GOES X-ray bandpass. On the other hand, solar observations have revealed that only about 1 per cent of the total energy of a flare is radiated in the GOES X-ray band (see \citeads{2006JGRA..11110S14W}, \citeads{2011A&A...530A..84K}). These two results imply that CMEs and flares in the Sun roughly follow $K^{\rm CME} \simeq 2\,E^{\rm FL}$, in agreement with the findings of \citetads{2012ApJ...759...71E}. This relation is indicated by the dashed line in Fig.~\ref{fig_8}. At face value, it would seem like the behavior of the flare-CME energy equipartition is not heavily altered by the presence of the large-scale field. We do not intend to imply that the solar relation is a good representation of our simulated events, and point out that their deviations from this relation do approach an order of magnitude. The escaping events also show a significantly different slope. However, the solar events on which the relation is based have all escaped from regions of the solar corona with a wide range of overlying magnetic field configurations (provided by the combination of the large-scale dipolar field and overlying fields anchored to the ARs) and into different ambient wind conditions, all of which must have some modulating influence on their speeds. In this sense, our simulations of escaping events do represent, in the coarsest way, an extrapolation into strong field conditions, such that the absence of much greater deviations from the mean solar relation is perhaps not surprising.

Finally, the status of our simulations (determined by visual inspection and the escape velocity of the perturbation) seems to closely match the position of the events relative to this empirical relation, with confined cases showing $K_{\rm C}^{\rm CME} < 2\,E^{\rm FL}$, and escaping CMEs following $K_{\rm E}^{\rm CME} \ge 2\,E^{\rm FL}$. Likewise, this correspondence serves to fill in the gap where the confined/escaping transition occurs, allowing us to locate the suppression threshold imposed by the overlying field. In this way, a $75$ G dipole acting as the large-scale component of the stellar field, is able to suppress CMEs with kinetic energies below $\sim$\,$3 \times 10^{32}$ erg. We stress that these results have been obtained under very specific conditions (see Sect.~\ref{sec_models}), and thus their generalization to other cases requires further investigation.  

\section{Summary and Conclusions}\label{sec_summary}

\noindent We considered a set of 3D MHD numerical simulations to study the suppression of coronal mass ejections (CMEs) by a large-scale magnetic field. This mechanism is expected to play an important role in determining the contribution from these eruptive phenomena to the mass loss and magnetic energy budget in active stars. The stellar wind and CME models considered here constitute the latest tools currently used for space weather prediction in the solar system.  

Guided by previous numerical studies, we showed that eruptions driven with the same parameters used to simulate strong CMEs on the Sun, can be completely confined by a large-scale surface magnetic field composed by a $75$~G dipole (aligned with the rotation axis of the star). We were able to put these results in the context of eruptive solar flares, leading to full suppression by this large-scale field of CME events associated with flares up to $\sim$X20 in the GOES classification.

The parameter space explored in our simulations included sufficiently strong events to escape the confining conditions. However, regardless of the suppressed or escaping state of the eruption, we found that the overlying field drastically reduced the CME speeds in comparison with expectations from solar observations and their extrapolations. On the other hand, the mass perturbed during the eruption was fairly consistent with the solar data, indicating a weaker influence on this parameter by the large-scale field. Still, our simulations indicate that CME interactions with the overlying field and the ambient stellar wind could lead to important structural changes of the eruption, such as partial confinement and CME fragmentation.        

Finally, our analysis revealed that only CMEs with kinetic energies greater than $\sim$\,$3 \times 10^{32}$ erg, would be able to escape the magnetic confinement imposed by the $75$~G dipolar large-scale field. For eruptions following the observed solar flare-CME behaviour, this could occur during flaring events with energies greater than $6 \times 10^{32}$~erg (GOES class $\sim$X70). Active stars not only display larger flare energies, but also stronger magnetic fields on small- and large-scales which, respectively, influence the generation and confinement of the CMEs. For this reason, a non-trivial extension of the results here presented is expected for those cases. Further numerical work will be pursued in order to determine the relative impact from additional factors on this suppression mechanism. This includes different stellar properties (e.g., mass, surface gravity, rotation period), large-scale field strength and complexity, characteristics of the small-scale field, influence from the mass-loaded filament, and possible CME-stellar wind interactions. 

\begin{acknowledgements}

\noindent We would like to thank the referee for his constructive comments which helped to improve the quality of this paper. Special thanks to Meng Jin for the very useful feedback during the initial stages of this project. JDAG was supported by Chandra grants AR4-15000X and GO5-16021X. JJD was funded by NASA contract NAS8-03060 to the Chandra X-ray Center and thanks the director, Belinda Wilkes, for continuing advice and support. OC and SPM were supported by NASA Living with a Star grant number NNX16AC11G. CG was funded by Chandra grants TM6-17001B and GO5-16021X. This work was carried out using the SWMF/BATSRUS tools developed at The University of Michigan Center for Space Environment Modeling (CSEM) and made available through the NASA Community Coordinated Modeling Center (CCMC). We acknowledge the support by the DFG Cluster of Excellence "Origin and Structure of the Universe". The simulations have been carried out on the computing facilities of the Computational Center for Particle and Astrophysics (C2PAP).

\end{acknowledgements}

\bibliographystyle{aasjournal}
\bibliography{Biblio}

\end{document}